\documentclass[prx,   notitlepage,superscriptaddress, nofootinbib]{revtex4-1}

\usepackage{fancyhdr}
\fancyheadoffset[L,R]{\headrulewidth}
\renewcommand{\headrulewidth}{0.6pt}

\usepackage{comment}

\usepackage{cancel}
\usepackage[normalem]{ulem}

\usepackage{cancel}

\usepackage{graphicx}
\usepackage{tikz,pgf}
\usepackage{color}
\usepackage{array}
\usepackage{xcolor}
\usepackage[dvips]{epsfig}
\usepackage{epsfig}
\usepackage{enumerate}
\usepackage[latin1]{inputenc}
\usepackage[T1]{fontenc}

\usepackage[hang]{subfigure}
\usepackage{bbm, dsfont}
\usepackage{amsfonts,amsmath,amssymb,stmaryrd}
\usepackage{ulem}

\usepackage{mathrsfs}
\usepackage{bbold}
\usepackage{setspace}

\usepackage{simplewick}

\newcommand{\bra}[1]{\langle #1 | \,}
\newcommand{\ket}[1]{\, | #1 \rangle}

\newcommand{\ga}{\ga}

\newcommand{\bl}{\begin{linenomath*}}
\newcommand{\el}{\end{linenomath*}}

\newcommand{\B}{\hat\Phi}

\newcommand{\Bd}{\hat\Phi^\dagger}

\newcommand{\bea}{\begin{eqnarray}}
\newcommand{\eea}{\end{eqnarray}}
\renewcommand{\vec}[1]{\mathbf{#1}}

\renewcommand{\a}{\hat a}
\newcommand{\ad}{\hat a^\dagger}

\newcommand{\be}{\hat b}
\newcommand{\bed}{\hat b^\dagger}
\renewcommand{\ga}{\hat\gamma}

\newcommand{\n}{\hat n}

\newcommand{\bd}{\hat b^\dagger}

\newcommand{\veck}{\mathbf k}

\newcommand{\vecp}{\mathbf p}

\newcommand{\vecr}{\mathbf r}

\definecolor{dgreen}{rgb}{0.0, 0.5, 0.0}

\usepackage{hyperref}

\begin{document}

\title{Molecular impurities interacting with a many-particle environment:\\[3pt] from  helium droplets to ultracold gases}

\author{Mikhail Lemeshko} 
\email{mikhail.lemeshko@ist.ac.at}
\affiliation{IST Austria (Institute of Science and Technology Austria), Am Campus 1, 3400 Klosterneuburg, Austria}

\author{Richard Schmidt} 
\email{richard.schmidt@cfa.harvard.edu}
\affiliation{ITAMP, Harvard-Smithsonian Center for Astrophysics, 60 Garden Street, Cambridge, MA 02138, USA}%
\affiliation{Physics Department, Harvard University, 17 Oxford Street, Cambridge, MA 02138, USA} %

\begin{abstract}

In several settings of physics and chemistry  one has to deal with molecules interacting with some kind of an external environment, be it a gas, a solution, or a crystal surface. Understanding molecular processes in the presence of such a many-particle bath is inherently challenging, and usually requires large-scale numerical computations. Here, we present an alternative approach to the problem -- that based on the notion of the angulon quasiparticle. We show that molecules rotating inside superfluid helium nanodroplets and Bose-Einstein Condensates form angulons, and therefore can be described by straightforward solutions of a simple microscopic Hamiltonian. Casting the problem in the language of angulons allows not only to tremendously simplify it, but also to gain insights into the origins of the observed phenomena and to make predictions for future experimental studies.

\end{abstract}

\maketitle
\vspace{-1.2cm}
\tableofcontents

\section{Introduction}

The properties of polyatomic systems we encounter in physics and chemistry can be extremely challenging to understand. First of all, many of these systems are strongly correlated, in the sense that their complex behavior cannot be easily deduced from the properties of their individual constituents  -- isolated atoms and molecules. Second, in realistic experiments these systems are usually found far from  their thermal equilibrium, as they are perturbed by the surrounding environment, be it a solution, a gas, or lattice vibrations in a crystal. Quite often, however, insight  into the behavior of such  complex many-body systems can be obtained from studying  the simplified problem of a single quantum particle  coupled to   an environment. These so-called `impurity problems' represent an important part of modern condensed matter physics~\cite{WeissBook, Breuer2002}.

\begin{figure}[t]
  \centering
 \includegraphics[width=0.9\linewidth]{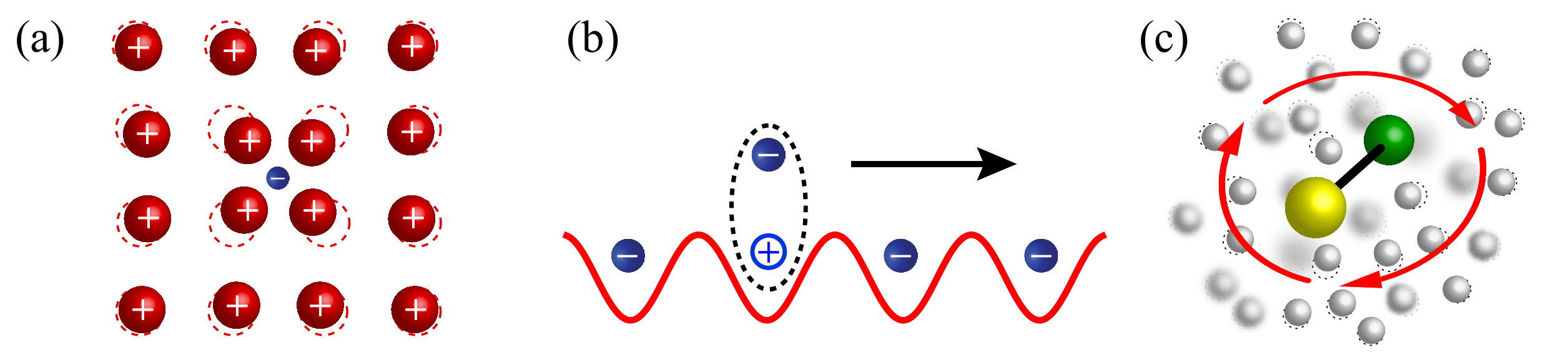}
  \caption{\label{quasi} Examples of quasiparticles. (a) Polaron: an electron  in a solid interacts with the crystal lattice. As a result, it is `dressed' by the resulting polarization cloud of nuclear displacements, and forms the so-called polaron. (b) Exciton: a bound state between an excited electron and a hole, which propagates in a semiconductor.  (c) Angulon: a quantum rotor dressed by a field of phonons. }
 \end{figure}

Interest in quantum impurities goes back to the classic works of Landau, Pekar, Fr\"ohlich, and Feynman, who studied motion of electrons in  crystals~\cite{LandauPolaron, LandauPekarJETP48, FrohlichAdvPhys54, FeynmanPR55, AppelPolarons, Devreese13}. In its most general formulation, such a problem involves the coordinates and momenta of all the electrons and nuclei in the crystal -- some $10^{23}$ degrees of freedom -- and is therefore intractable by any existing numerical technique. The problem, however, can be drastically simplified by using a trick very common among condensed matter physicists --  that of introducing  `quasiparticles.' A quasiparticle is a collective object, whose properties are qualitatively similar to those of free particles, however they quantitatively depend on the coupling between the particle and the environment. Fig.~\ref{quasi} shows a few examples of quasiparticles.
 For example, the behavior of an electron interacting with a crystalline lattice can be understood in terms of a so-called polaron quasiparticle, composed of an electron dressed by a coat of lattice excitations~\cite{EminPolarons, PolaronsExcitons}. A polaron effectively behaves as a free electron with a larger effective mass, whose exact magnitude depends on the value of the electron-lattice coupling. Casting the many-body problem in terms of polarons allowed to obtain insights into the physics of semiconductors and polymers~\cite{AppelPolarons,EminPolarons, PolaronsExcitons}, high-temperature superconductors~\cite{PolaronsHighTc}, and  a variety of other strongly correlated electron materials~\cite{Nagaev1975,Trugman1988,Aleksandrov_polaron_book},  $^3$He atoms immersed in superfluid $^4$He \cite{Bardeen1967}, and nuclear matter \cite{Forbes2014}. Furthermore, using `dressed impurities'  as a building block of a many-particle system is instrumental in several computational techniques, such as dynamical mean-field theory~\cite{Metzner1989,Georges1996,GullRMP11} and the dual boson approach~\cite{Rubtsov12}. 

Most of the impurity problems arose in the context of condensed-matter physics and were originally developed to treat point-like particles, such as electrons or   single spins. However, many of the impurity models can be successfully applied to more complex, composite quantum objects, such as atoms. For instance, various polaron models have been realized in the laboratory by immersing a single atom or ion in an ultracold Bose or Fermi gas~\cite{ChikkaturPRL00, SchirotzekPRL09, PalzerPRL09, KohstallNature12, KoschorreckNature12, SpethmannPRL12, FukuharaNatPhys13, ScellePRL13, Cetina15, MassignanRPP14,Jorgensen2016, Hu16, Cetina2016}. There, none of the electronically excited states of the atom can be populated both  due to weak interactions with the surrounding bath  and the small collisional energies involved. Therefore, the atom resides in its (usually spherically symmetric) ground state, and can be considered a point-like particle for all practical purposes. Due attention has also been paid to the spin degrees of freedom, which play a crucial role in the properties of  crystalline and amorphous solids~\cite{ChaikinLubensky}. For instance,  extensive research has been done on  single localized spins coupled to a bath of bosons~\cite{LeggettRMP87}, fermions~\cite{Anderson1961,LutchynPRB08,KnapPRX12}, and other spins~\cite{ProkofievSpinBath00}. However, although nonzero spin provides an impurity with an additional degree of freedom, in most cases discussed in the context of condensed matter physics they can still be described as entities with no extended spatial structure. For instance, in ultracold gases, the spin variable can be efficiently mapped onto the hyperfine states of the atoms or ions~\cite{BlochRMP08}, preserving the point-like nature of the latter.

Things change quite drastically, however, if one considers polyatomic impurities, such as molecules. Contrary to electrons or closed-shell atoms, molecules are extended objects which  possess a number of fundamentally different types of internal motion. The latter correspond to the rotational and vibrational degrees of freedom, which can couple to each other, as well as to the orbital and spin angular momentum of electrons, resulting in an involved energy level structure~\cite{KreStwFrieColdMol, LemKreDoyKais13, LevebvreBrionField2}. These `extra' degrees of freedom were key in numerous applications of molecular physics, from the classic example of the ammonia maser~\cite{GordonPR55}, to controlling the stereodynamics of chemical reactions~\cite{deMirandaNatPhys11}, to the recent measurements of the electron's electric dipole moment~\cite{ACMEedm}. Finally, these degrees of freedom occupy the low-energy part of the energy spectrum. As a result, they can be easily altered by the interactions with the surrounding medium, which makes them a rich resource to study novel aspects of many-body physics.

In several experimental realizations, isolated molecules are coupled to a bath of some kind. Even in the most inert environments, the lowest-energy degrees of freedom, such as molecular rotation, are perturbed by the reservoir. As an example, spectroscopists actively study molecules trapped inside inert matrices~\cite{MatrixIsolationBook}. However, even a weakly-polarizable crystalline matrix -- such as molecular parahydrogen~\cite{MomoseVibSpec04} -- splits the molecular states by its crystal field and causes rotational line broadening due to the molecule-phonon coupling. On the other hand, molecules are routinely trapped inside small droplets of superfluid helium~\cite{ToenniesAngChem04, StienkemeierJPB06, SzalewiczIRPC08}. This allows one to isolate a molecule inside a cryogenic environment and thereby perform accurate spectroscopic measurements, free from Doppler and collisional shifts. Furthermore, trapping single molecules in helium droplets makes it easier to study species that are reactive in the gas phase, such as free radicals~\cite{Kupper02}. While superfluid helium is the softest available matrix, it still induces changes in the rotational spectrum of the trapped species, such as the renormalization of the molecular rotational constant~\cite{ToenniesAngChem04}.
In the context of cold and ultracold gases, recent experimental progress allows to tame both translational and internal degrees of freedom by making use of electric, magnetic, and optical fields~\cite{LemKreDoyKais13, KreStwFrieColdMol, JinYeCRev12}.  In this way, experiments with cold controllable molecules open up a prospect to study their interaction with an environment in great detail, as it is currently being done with atomic impurities~\cite{ChikkaturPRL00, SchirotzekPRL09, PalzerPRL09, KohstallNature12, KoschorreckNature12, SpethmannPRL12, FukuharaNatPhys13, ScellePRL13, Cetina15, MassignanRPP14,Jorgensen2016,Cetina2016}. 

The goal of this tutorial is two-fold. On the one hand, we would like to introduce the reader to the impurity problem involving molecules, and put it into the context of the other, previously studied impurity problems~\cite{WeissBook}. On the other hand,  we aim to address at the same time  the physical chemists, working with molecules in helium nanodroplets, atomic physicists, interested in ultracold molecules, as well as condensed matter physicists, studying the polaron problem.
 
The tutorial is organized as follows. In Sec.~\ref{sec:superfluid} we introduce the concepts of superfluidity and Bose-Einstein condensation, in the context of liquid helium as well as ultracold atomic gases. Next, in Sec.~\ref{sec:ImpInSuperfluids} we describe recent advances on trapping molecules in superfluid helium droplets, and theoretical understanding of molecule-helium interactions. Over the years  there have been several extensive reviews describing the experimental techniques to trap molecules inside helium droplets~\cite{ToenniesAngChem04}, spectroscopy and dynamics of molecules~\cite{StienkemeierJPB06}, ionisation experiments~\cite{MudrichIRPC14}, as well as on the theoretical approaches to molecules in helium droplets~\cite{SzalewiczIRPC08}. Therefore, we will provide only a general survey of the topic.  In addition, in this tutorial we describe the experimental settings which can be employed to study molecular impurity physics with ultracold gases.

Sec.~\ref{sec:impurity} focuses on the main topic of   this tutorial -- a molecule coupled to a many-particle bath. We start from a  general microscopic Hamiltonian describing a rotating impurity coupled to a reservoir of bosons. We provide details on the rotational structures of rigid molecules and the angular momentum algebra involved. In order to describe bosons, we introduce the Bogoliubov approximation and transformation and discuss the origin of the roton minimum in superfluid helium. Finally, we derive the interaction between a molecule and a many-body bath from first principles. The goal of this tutorial is to derive the angulon Hamiltonian, Eq.~\eqref{Hamil1}, which is the central object of this tutorial. Next, in Sec.~\ref{sec:angulon} we study the Hamiltonian -- first, using perturbative and diagrammatic techniques. We show that even a simple model can describe rotational constant renormalization for molecules in helium droplets, and provide it with a transparent physical interpretation. Furthermore, we demonstrate that coupling of a molecule to a many-particle bath leads to the emergence of a novel  fine structure in the absorption spectrum, induced by many-body interactions. Finally, we describe a novel canonical transformation that drastically simplifies the solution in the limit of a slowly-rotating molecule. The main content of this tutorial closes with conclusions and an outlook, given in Sec.~\ref{sec:conclusions}. A short note on the relation between angular momentum operators in the laboratory and molecular frames of reference is given in Appendix A.

\section{Superfluidity and Bose-Einstein Condensation}
\label{sec:superfluid}

\begin{figure}[b]
  \centering
 \includegraphics[width=0.4\linewidth]{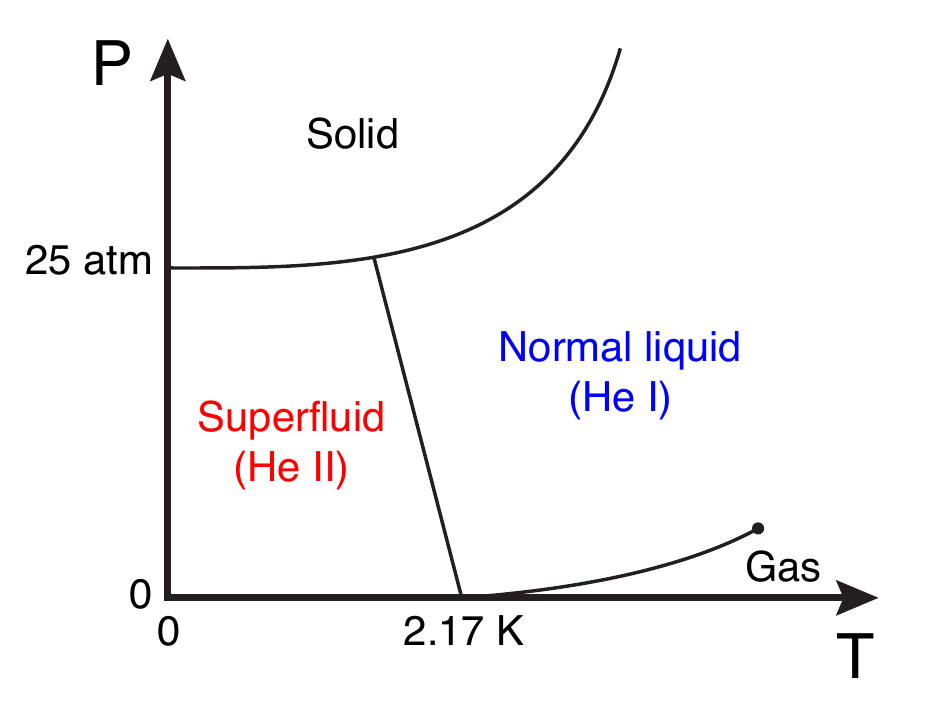}
  \caption{\label{Hediag} Schematic phase diagram of $^4$He, as a function of pressure, $P$, and temperature, $T$.}
 \end{figure}

Already in 1908, Kamerlingh Onnes from the University of Leiden observed that below 4.2 Kelvin helium gas turns into a low-density, colourless liquid. It took around a decade, however, to discover another, way more exotic phase transition which occurs in the vicinity of 2.17 Kelvin. In 1924, Kamerlingh Onnes and coworkers noticed  a  density change around the transition point~\cite{GriffinJPC09}. Within the next eight years, Keesom introduced the labels He I and He II for the two phases in his paper with Wolfke~\cite{KeesomPRAA28}, and measured the celebrated $\lambda$-shaped peak in the specific heat in collaboration with Clusius~\cite{KeesomPRAA32}. A schematic phase diagram of helium-4 is shown in Fig.~\ref{Hediag}. It is interesting to note that around that time several people observed a strange behavior of helium below 2.17 K~\cite{GriffinJPC09}, however none of them considered it important enough to describe it  in print. Apparently, around 1930, Keesom himself pointed out the experimental complications arising from the fact that He II easily leaks out through tiny holes in the apparatus~\cite{GriffinJPC09}. 
Only in late 30's were the hydrodynamic properties of He II properly measured by Allen and Misener~\cite{AllenNat38} in Cambridge, UK and, independently, by Kapitza~\cite{KapitzaNat38} in Moscow. They revealed that the low-temperature phase of helium features `superfluidity' (a term coined by Kapitza by analogy with superconductivity), i.e.,  it  is capable of moving through thin capillaries without any friction. The results were published in two back-to-back papers in Nature in 1938~\cite{AllenNat38, KapitzaNat38}, and thereby  pioneered the entire field of quantum liquids and solids~\cite{GriffinJPC09, BalibarLowTemp07, SchmittSuperfluidity, LeggettRMP99}.  

 Very soon after the experimental discovery, Fritz London  suggested~\cite{London38} that superfluidity is  closely related to Bose-Einstein condensation (BEC), whose  theoretical  fundamentals were known by that time \cite{Bose24, Einstein24}. After several discussions with London, Tisza developed the idea further and introduced the `two-fluid model,' postulating that two independent components are present in He II. The first, the  superfluid  component, represents a Bose condensate of the atoms where a single-particle quantum state is occupied by a macroscopic particle number. Tisza assumed that the resulting macroscopic coherence allowed for flux without friction or viscosity. The second, normal component, whose fraction depends on temperature, behaves as a regular viscous fluid~\cite{Tisza38}. Three years later, Landau introduced his version of the two-fluid model by quantizing the hydrodynamic theory of a classical liquid. His theory was phenomenological and did not make direct use of the Bose statistics of $^4$He particles. Moreover, in the introduction to his seminal paper~\cite{LandauPR41} Landau bluntly disagrees with Tisza and London, saying: ``Tisza's well-known attempt to consider helium II as a degenerate Bose gas cannot be accepted as satisfactory -- even putting aside the fact that liquid helium is not an ideal gas, nothing could prevent the atoms in the normal state from colliding with the excited atoms; i.e., when moving through the liquid they would experience friction and there would be no superfluidity at all.'' It took several decades to unify the theory of Landau with the BEC ideas of London and Tisza. After the first developments of Bogoliubov~\cite{Bogoliubov47} and Beliaev~\cite{Beliaev58}, several many-body calculations performed  between 1957 and 1964 allowed to reveal that the  superfluidity is indeed accompanied by Bose condensation of the helium atoms~\cite{Griffin99}. Nevertheless, measuring the Bose condensation in experiments with helium has proven challenging, and only indirect techniques such as neutron scattering have been successfully used~\cite{GriffinJPC09, BalibarLowTemp07, SchmittSuperfluidity, LeggettRMP99}.
 
 A direct proof of the phase transition  to a Bose-Einstein condensate  was  achieved  only in 1995 in experiments with dilute alkali gases~\cite{Pitaevskii2016, PethickSmith}. As opposed to experiments with helium, in ultracold gases BEC can be detected easier than superfluidity, which was observed later as well~\cite{OnofrioPRL00}. A   flow without friction is not the only peculiar property of the superfluids. For instance, if rotated, superfluids develop vortices -- tiny strings carrying quantized angular momentum, whose number grows with the speed of rotation. Vortices, as fingerprints of the superfluid phase, have also been detected in numerous experiments on helium, as well as on ultracold alkali gases~\cite{Pitaevskii2016, PethickSmith, GomezPRL12, Gomez906}.
 
 Since the phenomenological model of Landau described all experimentally observed properties of superfluid helium, the ideas of London and Tisza remained in the shade for several decades. However, one can speculate that if the BEC was observed in ultracold gases earlier than superfluidity in helium, the validity of the London-Tisza theory would have never been in question. Since in this tutorial we did not provide theoretical details on the theory of superfluidity or BEC, the interested reader is referred to several great books that have been published on the subject~\cite{Pitaevskii2016, LeggettQuantLiquids, PinesNozieres}.

Let us now discuss the conditions necessary to observe superfluidity and BEC. Since these are inherently quantum effects, they manifest themselves at large values of the de Broglie wavelength,
\begin{equation}
\lambda = \frac{h}{p}
\end{equation}
Here $h$ is Planck's constant and $p$ is the particle's momentum. For particles of mass $m$ at temperature $T$, the momentum $p \sim (m k_B T)^{1/2}$, where $k_B$ is Boltzmann's constant. For matter at `normal' conditions, the resulting de~Broglie wavelength is much smaller than the interparticle distance. As a result, the properties of gaseous or liquid matter can be successfully described by laws of classical statistical physics, treating individual atoms and molecules as rigid spheres or ellipsoids. However, this is not the case in the regime of low temperatures or high densities, where the average distance between the particles, $\langle r \rangle$, becomes shorter than the de~Broglie wavelength, $\langle r \rangle \lesssim \lambda$. Given that $\langle r \rangle$ can be expressed through the particle number density, $n$, as $\langle r \rangle = n^{-1/3}$, we obtain the following relation:
\begin{equation}
\label{qm_effects}
k_B T \lesssim n^{2/3} \hbar^2/m
\end{equation}
From Eq.~\eqref{qm_effects} one can see that the role played by quantum effects in the properties of a gas or liquid is determined by the  interplay between its density and temperature. In dilute alkali gases, particle densities are usually on the order of $10^{14} - 10^{15}$ particles per cm$^{3}$. Therefore, in order to achieve BEC, one needs to cool the gas down to microKelvin temperatures. In superfluid helium, on the other hand, the densities are much larger  and on the order of  $10^{22}$ cm$^{-3}$~\cite{Donnelly1998}. As a result, the BEC transition occurs at a relatively high temperature of $\sim 2$ Kelvin. One should keep in mind that Eq.~\eqref{qm_effects} does not take interatomic interactions into account and therefore provides  only  a  rough  estimate  for the transition temperature. In dilute alkali gases, however, the interactions between the atoms are sufficiently weak such that almost all of them can reach the BEC state. On the other hand, due to strong He--He interactions, only a small fraction of superfluid helium ($\sim 6-8$\%) reaches the BEC state at $T=0$.

Out of all bosonic atoms in the periodic table,  $^4$He is the only one that becomes superfluid `naturally,' i.e.\ upon a transition from a normal liquid phase. What is so special about helium? The reason lies in the interplay between the kinetic energy of the atoms and their mutual interactions. Helium is extremely light, and only weakly polarizable, and therefore the kinetic energy (`zero-point motion') of the helium atoms is always larger than the interactions between them. This precludes the atoms from freezing into a crystalline lattice, unless an external  pressure is applied. Other elements, on the other hand, solidify at much higher temperatures compared to those given by Eq.~\eqref{qm_effects}, and the superfluid transition never takes place.

While Eq.~\eqref{qm_effects} provides a criterion concerning the importance of the quantum mechanical effects, additional phenomena arise from quantum statistics. There, the particles are divided into bosons, whose collective wavefunction stays intact after particle exchange, and fermions, whose collective  wavefunction changes sign. In order for quantum statistics to play a role, however, the particles are required to be able to physically change places. This is the case in gases and liquids, but not in solids where tunnelling of the nuclei building the lattice is greatly suppressed. Therefore, quantum statistics  plays no role for the lattice sites in solids.  The situation is different in molecular physics. Here indistinguishable nuclei can exchange position and, hence, quantum statistics  becomes relevant.  For instance, if one considers a molecule, composed of two bosonic atoms, e.g. $^{12}$C$_2$, its wavefunction has to be symmetric under exchange of the nuclei. As a result, odd rotational states will be missing, since they would break the symmetry. If one replaces one of the nuclei by its isotope, say $^{14}$C, so that the nuclei become distinguishable, all rotational states will show up in the spectrum~\cite{FurtenbacherAstrJ16}.

\section{Molecular impurities trapped inside superfluids}
\label{sec:ImpInSuperfluids}

\begin{figure}[b]
  \centering
 \includegraphics[width=0.5\linewidth]{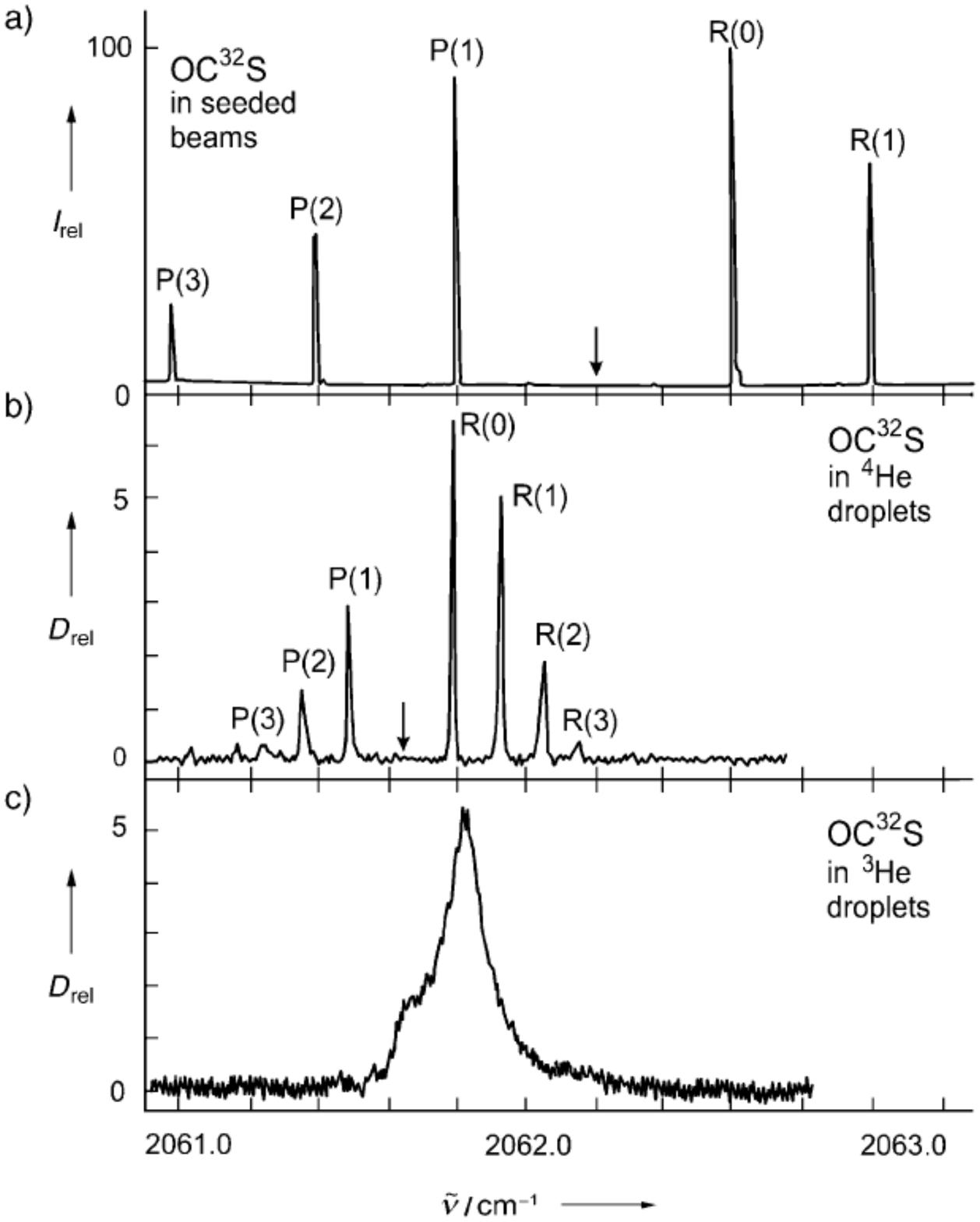}
  \caption{\label{OCS}  Rovibrational spectra of the OCS molecule. (a) In the gas phase. (b) Inside superfluid $^4$He: the lines are sharp, but the rotational level splittings are renormalized. (c) Inside non-superfluid $^3$He: the rotational lines are completely broadened by the helium environment.  Adapted with permission from Ref.~\cite{ToenniesAngChem04}. }
 \end{figure}
 
For several decades after the discovery of superfluidity, only macroscopic, hydrodynamic properties of helium have been studied.  Employing \textit{microscopic} probes, on the other hand, seemed extremely challenging, since it turned out that superfluid helium is averse to mixing with impurities. Only in the 1990s it was demonstrated that atoms and molecules  can be trapped in helium if the latter forms little droplets~\cite{ToenniesAngChem04, SzalewiczIRPC08}. Over the following years, trapping atoms, molecules, and ions inside the superfluid helium droplets -- sometimes called `nanocryostats' -- emerged as an important tool in spectroscopy~\cite{ToenniesAngChem04}. Superfluid helium droplets cool the molecules to $\sim 0.4$ Kelvin~\cite{StienkemeierJPB06} and isolate them from external perturbations. This allows to record spectra free of collisional and Doppler broadening, as well as to trap  and study species reactive in the gas phase. By now, matrix isolation spectroscopy based on helium droplets evolved into a large field, which has been the subject of several review articles~\cite{ToenniesARPC98, ToenniesAngChem04, SzalewiczIRPC08, StienkemeierJPB06, MudrichIRPC14}. Therefore, here we describe only the main effects arising due to the interactions between the molecule and the helium droplet, without providing too many technical details.

  \begin{table}[b]
\begin{tabular}{| c | c | c | c | c| }
  \hline			
  Molecule & $B$ & $B^\ast$ & $B^\ast/B$ & Ref. \\
  \hline
  HF & 19.787 & 19.47 & 0.98 & \cite{NautaJCP00} \\
  HCN & 1.478 & 1.204 & 0.81 &  \cite{NautaPRL99, ConjusteauJCP00}  \\
 CO$_2$ & 0.39 & 0.154 & 0.39 & \cite{NautaJCP01} \\
  OCS & 0.2029 & 0.0732 & 0.36 & \cite{GrebenevJCP00} \\
 N$_2$O & 0.4187 & 0.0717 & 0.171 & \cite{NautaJCP01, XuPRL03}  \\
  \hline  
\end{tabular}
\caption{\label{tab:mol}  The rotational spectrum of molecules in superfluid helium droplets  can be approximated as $B^\ast J(J+1)$, where $B^\ast$ is the effective rotational constant, and $J=0,1,2 \dots$ labels the rotational levels. The table gives  examples of $B^\ast$ in comparison with the gas-phase value of the rotational constant, $B$, for linear rotor molecules. Energies are given in cm$^{-1}$. 
}
\end{table}
 
Among all inert gases conventionally used for matrix isolation~\cite{MatrixIsolationBook}, helium represents the `softest,' i.e.\ the least polarizable environment. Furthermore, since at atmospheric pressure helium does not form a crystal, it does not break the rotational symmetry of the molecules, unlike another weakly-interacting matrix -- molecular \textit{para}-hydrogen~\cite{MomoseVibSpec04}. As a result, the impurity molecules interact with the surrounding matrix only weakly.
Figure~\ref{OCS} shows the legendary experimental spectrum of the OCS molecule -- a prototypical linear rotor -- in the gas phase, in superfluid $^4$He droplets, as well as in droplets of $^3$He, not superfluid at this temperature. Comparing panels (a) and (b) one  can see that the only change induced by the $^4$He environment is a renormalized spacing between the transition lines. The overall structure of the spectrum remained the same, and the helium environment does not lead to any substantial broadening of the lines~\cite{ToenniesAngChem04}.

\begin{figure}[t]
  \centering
 \includegraphics[width=0.8\linewidth]{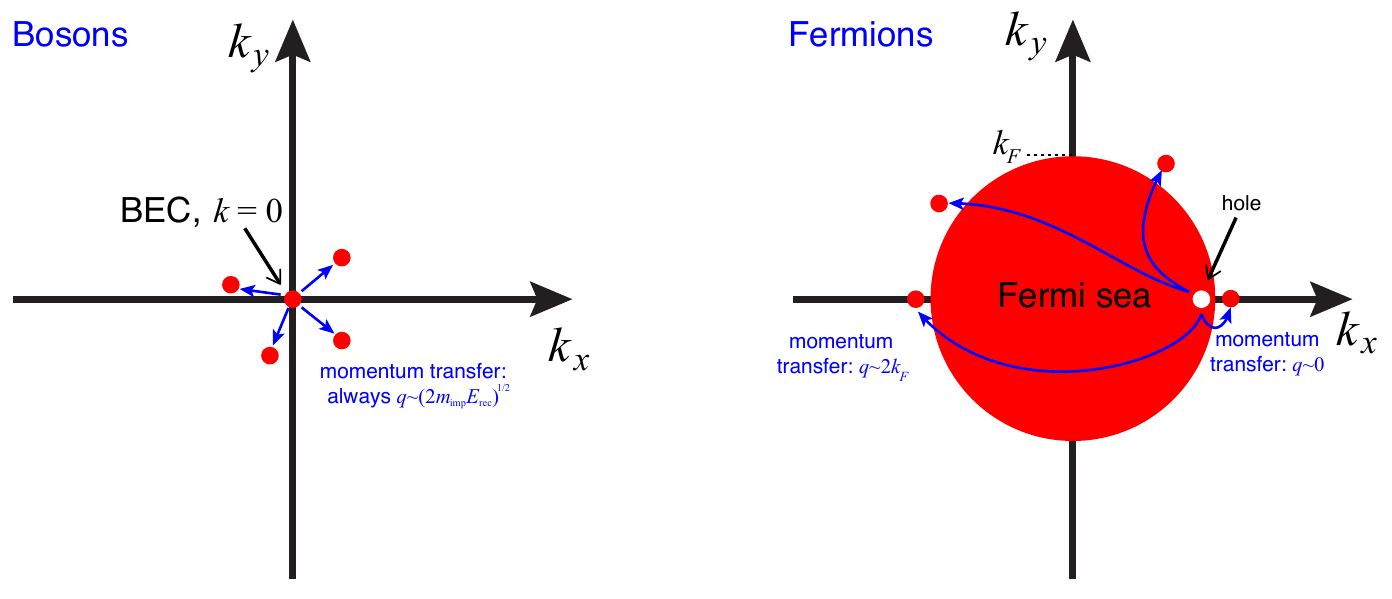}
  \caption{\label{BosonsFermions}   When particles in the medium scatter off the impurity, their momentum changes from $k_\text{in}$ to $k_\text{out}$, while the impurity experiences a recoil energy $E_\text{rec}=q^2/(2m_\text{imp})$ where $q=|\vec{k}_\text{in} - \vec{k}_\text{out}|$ is the momentum transfer. In the case of a BEC (left), all bosons are initially in the $k=0$ state so that the scattering  also leads to a kinetic energy cost for them. This is drastically different for fermions (right), where the same recoil energy of the impurity can create all kinds of excitations of the fermionic bath with momentum transfers ranging from $q \sim 0$ to $q \sim 2k_F$, with no additional energy cost.}
 \end{figure}

Panel (c), on the other hand, shows that interactions with a fermionic $^3$He environment broaden the spectral lines of the same molecule to the point that they are no longer resolved. In Ref.~\cite{GrebenevScience98}, where these spectra were reported first, such a behavior was attributed to the hydrodynamic properties of helium, i.e.\ the soft, non-disturbing nature of the superfluid phase. However, as was described later in Ref.~\cite{Babichenko99}, the reason rather lies in quantum statistics, which leads to a drastically different  phase space for available scattering processes in bosonic and fermionic environments, as schematically illustrated in Fig.~\ref{BosonsFermions}.

One can understand this effect as follows. When an impurity is immersed in a bath of atoms, the atoms scatter off the impurity, which leads to line broadening as well as  shifts of the spectral lines.  In each scattering event with the impurity, the atom's momentum is changed from $\vec{k}_\text{in}$  to $ \vec{k}_\text{out}$, which is accompanied by a transfer of momentum $\vec{q}=\vec{k}_\text{in}-\vec{k}_\text{out}$  to the impurity. If the bath consists of a Bose-Einstein condensate, most atoms are in a low kinetic energy state,  so that the incoming  momentum  is small, $\vec{k}_\text{in}  \approx 0$.  When colliding with the impurity,  atoms can be transferred into finite momentum states, $\vec{k}_\text{out}$. Since the bosons are initially in their kinetic ground state, their excitation always costs a finite amount of energy. Consequently,  such processes are suppressed.   In contrast, in the case of $^3\text{He}$ one deals with a degenerate Fermi gas where, due to Pauli blocking, all scattering states up to the Fermi momentum, $k_F$, are occupied, see the right panel of Fig.~\ref{BosonsFermions}. Hence the fermions which scatter off the impurity can start off a plethora of initial states  just below the Fermi surface, $|\veck_\text{in}| \lesssim k_F$, to unoccupied states just above the Fermi surface, $|\veck_\text{out}| \gtrsim k_F$. Due to the vector addition of momenta, the resulting momentum transfer, $q=|\vec{k}_\text{in}-\vec{k}_\text{out}|$, can range from $q=0$ to $q=2k_F$, whithout changing the  energy of the fermions in a substantial way. Consequently, there is a very large number of such scattering processes available, leading to much stronger effects on the observed spectral features as compared to the case of a BEC environment. Note that in this simplified argument we assumed that the mass of the impurity $m_\text{imp}$ is large compared to that of the fermions in the environment. In this case, the recoil energy $E_\text{rec} = q^2/(2m_\text{imp})$ gained by the impurity is small and can be neglected. As a result, the number of pathways leading to decoherence and renormalization of the rotational states is substantially larger in a fermionic bath compared to the bosonic one. 

Table~\ref{tab:mol} compares the rotational constants in the gas-phase with those inside superfluid helium for several linear-rotor molecules. We will discover later that this effect is not entirely due to the hindering of rotation by the presence of He atoms. The purpose of the present tutorial is to demonstrate that such a classical picture is by far incomplete. One can see that the  rotational constants of heavier molecules are  subject to stronger renormalization than the lighter ones. This comes from the fact that for slowly rotating molecules the anisotropic part of the molecule-helium  interaction is, in general,  more substantial compared to the kinetic energy. It is worth remembering that quite often it is said that the helium matrix perturbs the rotational states only weakly. While this is true concerning the widths of the rotational lines, the coupling to the environment is, strictly speaking, strong for all the molecules where $B^\ast /B \ll 1$, since in this case the interactions are much larger than the kinetic energy.

Theoretically, calculating the properties of superfluid helium (even without the impurities involved) represents a formidable challenge. While, compared to bulk helium, the droplets consist of a relatively `small'  number of atoms (several hundreds to millions), their properties largely coincide with the bulk properties. On the other hand, numerical calculations (e.g. those based on Monte-Carlo algorithms) become unfeasible already at the level of hundreds of atoms. Nevertheless, even calculations for small systems allow to understand the properties of  molecules inside helium nanodroplets.  This is in line with several studies showing that tens of helium atoms suffice to observe superfluidity~\cite{GrebenevScience98, SurinPRL08}.

Over the years,  the groups of Whaley, Zillich, Krotscheck, Blume, Roy, Moroni, and others performed sophisticated numerical calculations for several different molecules  trapped inside small and large He$_n$ clusters (with $n \lesssim 100$). A detailed review of their work is provided in Refs.~\cite{ToenniesAngChem04, SzalewiczIRPC08}. Such calculations  revealed the superfluid fraction of the helium droplet, the fraction of He atoms forming a BEC, as well as the spectrum of the molecular impurity, from which the change of its rotational constant.  Since, due to a relatively high density of superfluid helium, the helium atoms are densely packed in the vicinity of the molecule, the details of the two-body molecule-helium potentials are of particular importance. As a result, the accuracy of the potential energy surfaces plays a crucial role in numerical simulations. An excellent review of theoretical studies of molecules in helium nanodroplets was provided in Ref.~\cite{SzalewiczIRPC08}, therefore we do not describe the computational machinery of the methods in this tutorial. It is important to note, however, that while \textit{ab initio} quantum mechanical calculations are required to quantitatively reproduce the experiments on rovibrational spectroscopy, simpler -- preferentially analytic -- models are necessary in order to understand interactions of molecules with the surrounding environment. In the following sections we introduce such models, based on the angulon quasiparticle, and show that they provide insights into molecular rotation inside superfluid helium.

While studies of molecules in helium droplets traditionally belong to the domain of chemical physics and physical chemistry, during the last decades  physicists working with ultracold quantum gases have achieved an unprecedented control of internal and external atomic degrees of freedom~\cite{Pitaevskii2016, PethickSmith}. 
There, the physics of structureless impurities -- polarons -- has been extensively studied. As one example, an ion or atom immersed {in a sea of bosons or fermions} dresses  itself with  a `coat' of many-particle excitations, thereby turning into a Bose- or Fermi-polaron (depending on the bath statistics). {Both kinds of polarons} have been realized in cold atomic gases by a number of groups~\cite{ChikkaturPRL00, SchirotzekPRL09, PalzerPRL09, KohstallNature12, KoschorreckNature12, SpethmannPRL12, FukuharaNatPhys13, ScellePRL13, Cetina15, MassignanRPP14, Jorgensen2016,Cetina2016}.

 In recent years it became possible to create samples of cold molecules using magneto- and photo-association  in ultracold gases, laser and Sisyphus cooling, as well as Stark and Zeeman deceleration~\cite{LemKreDoyKais13, BasChemRev12, JinYeCRev12, KreStwFrieColdMol}. {It is within reach of state-of-the-art experiments} that these molecules can be selectively prepared in a single hyperfine state and immersed into a Bose or Fermi gas with controlled density and interactions.  In addition, one can apply external electric, magnetic, or laser fields to steer molecular rotation. These developments pave the way to studying coupling of molecular rotation with a many-particle setting in a fully-controlled manner, which we describe in the following section.

We note that during the last decade there have been numerous proposals on many-particle physics with ultracold molecules~\cite{LewensteinBook12, LemKreDoyKais13, JinYeCRev12, KreStwFrieColdMol}, some of which have already been realized in the laboratory~\cite{YanNature13}. {Most such proposals rely on dipole-dipole interactions, which manifest themselves when their magnitude is on the order of the molecular kinetic energy. One way to reach such a regime is to prepare a high-density molecular sample, in order to decrease the average intermolecular distance and thereby increase the dipole-dipole interactions. This is quite challenging to achieve, partly because of the reactive collisions between the particles. Another approach relies on lowering the sample's temperature in order to decrease the average kinetic energy of the molecules.  This is, in turn, challenging, since the molecular laser cooling techniques are quite limited while sympathetic and evaporative cooling is precluded by reactive collisions~\cite{LemKreDoyKais13}}. In the following sections we show that a single molecule coupled to an ultracold gas represents an elementary building block of a many-particle system, and thereby allows to study {novel physics, inaccessible with atomic impurities}. As a pleasant bonus, high densities of molecules, leading to unpleasant chemical reactions and loss of atoms, can be avoided, while many intriguing aspects of many-body impurity physics can still be probed.

\section{Theoretical description of the molecular impurity problem}
\label{sec:impurity}

In this section, we derive the Hamiltonian for a molecular impurity coupled to a bath of bosons, preparing the reader for the next section, where the problem is cast using the concept of quasiparticles. We start from  a first-principle Hamiltonian which describes the systems in terms of bosonic atoms interacting with a single molecule. In its most general form it is given by: 
\begin{equation}
\label{Hh}
\hat H =  \hat H_\text{mol} + \hat H_\text{bos} + \hat H_\text{mol-bos}
\end{equation}
where the three terms are, respectively, the Hamiltonians of the isolated molecule, the unperturbed bath of bosons, and the interaction between molecule and bosonic bath. In what follows we describe each term in detail.

\subsection{Molecular Hamiltonian}

In addition to the {electronic as well as fine and hyperfine structure} found in atoms, molecules possess additional types of internal motion, such as rotation and vibration. The characteristic energies for the electronic transitions correspond to visible and ultraviolet light (frequencies of $10^{14} - 10^{15}$~Hz, {wavelengths in the range of 300--3000~nm}),  while  vibrational excitations mostly lie in the infrared region ($10^{13} - 10^{14}$~Hz,  {wavelengths in the range of 3--30~$\mu$m}), and rotational transitions correspond to microwave frequencies ($10^{9} - 10^{11}$~Hz, {wavelengths in the range of 3--300~mm}). Usually, the natural lifetimes of the electronic states are very short (at most on the order of microseconds), and molecules reside in their ground electronic state. Furthermore, at temperatures {of $T\sim 1$~K}, as they are present inside superfluid helium and which correspond to {$\nu = k_B~T/h \sim 10^{10}$ Hz}, molecules are cooled to the ground vibrational state, with very few rotational states populated {in the case of small molecules}. Moreover, interactions with the environment are rarely strong enough to disturb the vibrational and electronic spectrum of the molecules. Therefore, within the Born-Oppenheimer approximation, we can focus exclusively on the rotational degrees of freedom and disregard other internal degrees of freedom, of both the molecule and the atoms in the environment.  Furthermore, here we will consider molecules whose translational degrees of freedom are frozen.  While in some situations the translational motion of the molecules might play a role, neglecting it is a good approximation both for molecules at ultracold temperatures~\cite{JinYeCRev12, KreStwFrieColdMol, LemKreDoyKais13} and molecules inside helium droplets~\cite{ToenniesAngChem04}.

Within these approximations, the low-energy sector of the molecular Hamiltonian can be expressed as:
\begin{equation}
\label{Hmol}
\hat H_\text{mol} = \hat H_\text{rot} + \hat H_\text{pert}
\end{equation}
where $\hat H_\text{rot}$ describes the rotations of a molecule as a rigid body, while $\hat H_\text{pert}$ describes various perturbations to the rotational spectrum. The latter mostly arise due to the spin-orbit, spin-rotation, and spin-spin interactions and depend on the particular electronic state the molecule is in~\cite{LevebvreBrionField2}. For simplicity, we will consider closed-shell ($^1\Sigma$) molecules where such perturbations are negligible.

The rotational Hamiltonian of a rigid rotor is obtained by quantizing its classical counterpart. In the most general case it is given by:
\begin{equation}
\label{Hrot1}
\hat H_\text{rot} = A \hat J_{x}'^2 + B \hat J_{y}'^2 + C \hat J_{z}'^2
\end{equation}
Here, the operators  $\hat J_{i}'$ define the projections of the angular momentum on the molecular-frame coordinate axes, {$(x, y, z)$}, chosen to coincide with molecular axes of symmetry.

The so-called rotational constants, $A, B$, and $C$,  are expressed through the corresponding moments of inertia, $I_i$, as
\begin{equation}
\label{ABC}
{A = \frac{1}{2I_{x}}, \hspace{0.5cm} B = \frac{1}{2I_{y}}, \hspace{0.5cm} C = \frac{1}{2I_{z}}}.
\end{equation}
As already done in this equation, in order to simplify formulae we set $\hbar \equiv 1$ from now onwards. In such a convention, momentum and energy have dimensions of reciprocal length and time, respectively. The special cases of $\hat H_\text{rot}$ cover  species of various geometry:  {$A=B; C=0$ corresponds to linear molecules (KRb, CO$_2$); $A=B=C$ corresponds to spherical tops (CCl$_4$, C$_{60}$); $A=B\neq C$ are called symmetric tops, where $A=B<C$ define prolate symmetric tops (CH$_3$Cl,  CH$_3$C$=$CH) and $A=B>C$ oblate symmetric tops (NH$_3$, C$_6$H$_6$, CHCl$_3$); and $A \neq B \neq C$ correspond to asymmetric tops (H$_2$O, CH$_2$Cl$_2$). In the following sections, we provide details on molecules with different rotational structure. Detailed descriptions of these cases can be found in Refs.~\cite{TownesSchawlow, BernathBook, LevebvreBrionField2}. First, however, let us discuss the properties of the rotational angular momentum operators.}

\subsubsection{Angular momentum operators in the molecular and laboratory frames}

{The molecular (or `body-fixed') coordinate system of Eq.~\eqref{Hrot1} is introduced in addition to the laboratory (or `space-fixed') one, whose axes we will label as $(X,Y,Z)$. Furthermore, in both coordinate systems one can introduce the spherical components of angular momentum operators as:
\begin{align}
\label{J0x}
	 \hat{J}_0' &= \hat{J}_z' \\
 \label{Jplusx}
	 \hat{J}_{+1}' &= -\frac{1}{\sqrt{2}} \left(\hat{J}_x' + i\hat{J}_y' \right)\\
\label{Jminusx}
	 \hat{J}_{-1}' &= \frac{1}{\sqrt{2}} \left(\hat{J}_x' - i\hat{J}_y' \right)
\end{align}
(see Appendix~\ref{sec:appendixAngular} for details).
Working in this representation, the molecular-frame components, $\hat J'_i$, of the angular momentum operator, $\hat{\mathbf{J}}$, can be expressed via the laboratory-frame components, $\hat J_k$, as:
\begin{equation}
\label{JiPrimeviaJi}
 	\hat{J}'_{i}   = \sum_k \hat D^{1}_{k, i}  (\hat \phi,  \hat \theta, \hat \gamma) \hat{J}_{k},
\end{equation}
where $i,k = \{-1, 0, +1\}$, $\hat D^{l}_{k, i} (\hat \phi,  \hat \theta, \hat \gamma)$ are so-called Wigner rotation matrices~\cite{VarshalovichAngMom}, and $(\hat \phi,  \hat \theta, \hat \gamma)$ give the Euler angles of the relative orientation of the molecular coordinate system with respect to the laboratory frame.  Note that the angles determining the instantaneous orientation of the molecular frame with respect to the laboratory frame are given by   operators that measure the orientation of the molecular impurity.  In order to clarify this point, let us consider the properties of these operators in more detail. }

{Upon acting on an eigenstate of the angles, $\ket{\phi, \theta, \gamma} $, the angle operators $(\hat \phi,  \hat \theta, \hat \gamma)$ are replaced by their eigenvalues
 \begin{align}
\label{AngleOp1}
\hat \phi \ket{\phi, \theta, \gamma} &=  \phi \ket{\phi, \theta, \gamma}\\
\hat \theta \ket{\phi, \theta, \gamma} &=  \theta \ket{\phi, \theta, \gamma} \\
\hat \gamma \ket{\phi, \theta, \gamma} &=  \gamma \ket{\phi, \theta, \gamma}
\end{align} 
Since the Wigner rotation matrices $\hat D^{l}_{k, i} (\hat \phi,  \hat \theta, \hat \gamma)$  are analytical functions of the angles, the following relation is satisfied:
 \begin{equation}
\label{DlkiAngle}
\hat D^{l}_{k, i}  (\hat \phi,  \hat \theta, \hat \gamma) \ket{\phi, \theta, \gamma} = D^{l}_{k, i}  (\phi,  \theta, \gamma) \ket{\phi, \theta, \gamma} 
\end{equation} 
Thus, by acting on a given molecular state, the $\hat D$-operator `measures' the instantaneous molecular orientation in the laboratory frame, and the corresponding projections $\hat J'_i$ can be evaluated using Eq.~\eqref{JiPrimeviaJi}.}

{One should keep in mind that the notions of `space-fixed' and `body-fixed' belong only to the projections of the angular momentum operator. That is, $\hat J_i'$ are not evaluated using the coordinate and momentum operators, $\hat{\mathbf{r}}$ and $\hat{\mathbf{p}}$, in the  molecular  frame, but give the projections of the laboratory-frame angular momentum $\hat{\mathbf{J}}$ onto the rotating axes of the molecule. The amount of angular momentum possessed by the molecule does not depend on the coordinate system  and it  can be easily verified that  $\hat{\mathbf{J}}^2 \equiv  \hat{\mathbf{J}}'^2$.
}

{There is one more important consequence of $(\hat \phi,  \hat \theta, \hat \gamma)$ being operators: the Wigner $\hat D$-matrix of Eq.~\eqref{JiPrimeviaJi} does not commute with the angular momentum operators $\hat{J}_{k}$ and $\hat{J}'_{i}$; for the respective commutation relations see Eqs.~\eqref{JiComm}--\eqref{JiPrimeCommAst} of Appendix~\ref{sec:appendixAngular}. Using these commutation relations together with Eq.~\eqref{JiPrimeviaJi} one uncovers a surprising property: the molecular-frame operators $\hat J'_i$ possess anomalous commutation relations~\cite{NautsAmJPhys10}. For example, while for the laboratory-frame components $[\hat J_X, \hat J_Y] = i \hat J_Z$, for the molecule-frame components $[\hat J'_x, \hat J'_y] = - i \hat J'_z$. This  unusual result can be understood in terms of time-reversal symmetry~\cite{ZareAngMom, JuddDiatomic}. If an observer rotates along with the molecular frame, from their point of view the laboratory frame spins in the opposite direction compared to the molecular rotation with respect to the laboratory frame. Rotation in an opposite direction corresponds to an inversion of time, $t \to -t$, which leaves the coordinates intact, $\mathbf{r} \to \mathbf{r}$, but changes the signs of momenta, $\mathbf{p} \to -\mathbf{p}$. Since $\mathbf{J} = \mathbf{r} \times \mathbf{p}$, the signs of angular momenta are also changed, $\mathbf{J} \to -\mathbf{J}$, which leads to the minus  sign in the commutation relations. Note that this is analogous to performing complex conjugation, i.e.
 the replacement of $i \to -i$, which corresponds to the time-reversal operation in quantum mechanics~\cite{SakuraiQM}.

\subsubsection{Linear molecules}

In the case of linear molecules, Eq.~\eqref{Hrot1} reduces to 
\begin{equation}
\label{Hrot2}
{\hat H_\text{rot} = B \hat J_{x}'^2 + B \hat J_{y}'^2 \equiv B \mathbf{\hat{J}^2}}
\end{equation}

Thus the quantum state of a rigid-rotor molecule is defined by the eigenvalues of $\mathbf{\hat{J}^2}$ and one of projections with respect to the laboratory coordinate axes, usually chosen to be $\hat{J}_Z$:
\begin{align}
\label{Jeigen}
 \mathbf{\hat{J}^2} \vert j, m \rangle &= j(j+1) \vert j, m \rangle\\ \notag
  \hat{J}_Z \vert j, m \rangle &= m \vert j, m \rangle
\end{align}
In the absence of external fields, the eigenstates of a rigid linear rotor form $(2j+1)$-fold degenerate multiplets with energies $E_j = B j(j+1)$.

Often it is convenient to work in the angular representation, where the linear rotor wavefunctions are given by spherical harmonics~\cite{VarshalovichAngMom, ZareAngMom, BiedenharnAngMom}:
\begin{equation}
\label{JYlm}
\langle \theta, \phi \vert j, m \rangle = Y_{j m} (\theta, \phi)
\end{equation}

\subsubsection{Symmetric-top molecules}
\label{sec:symtops}

\begin{figure}[t]
  \centering
 \includegraphics[width=0.5\linewidth]{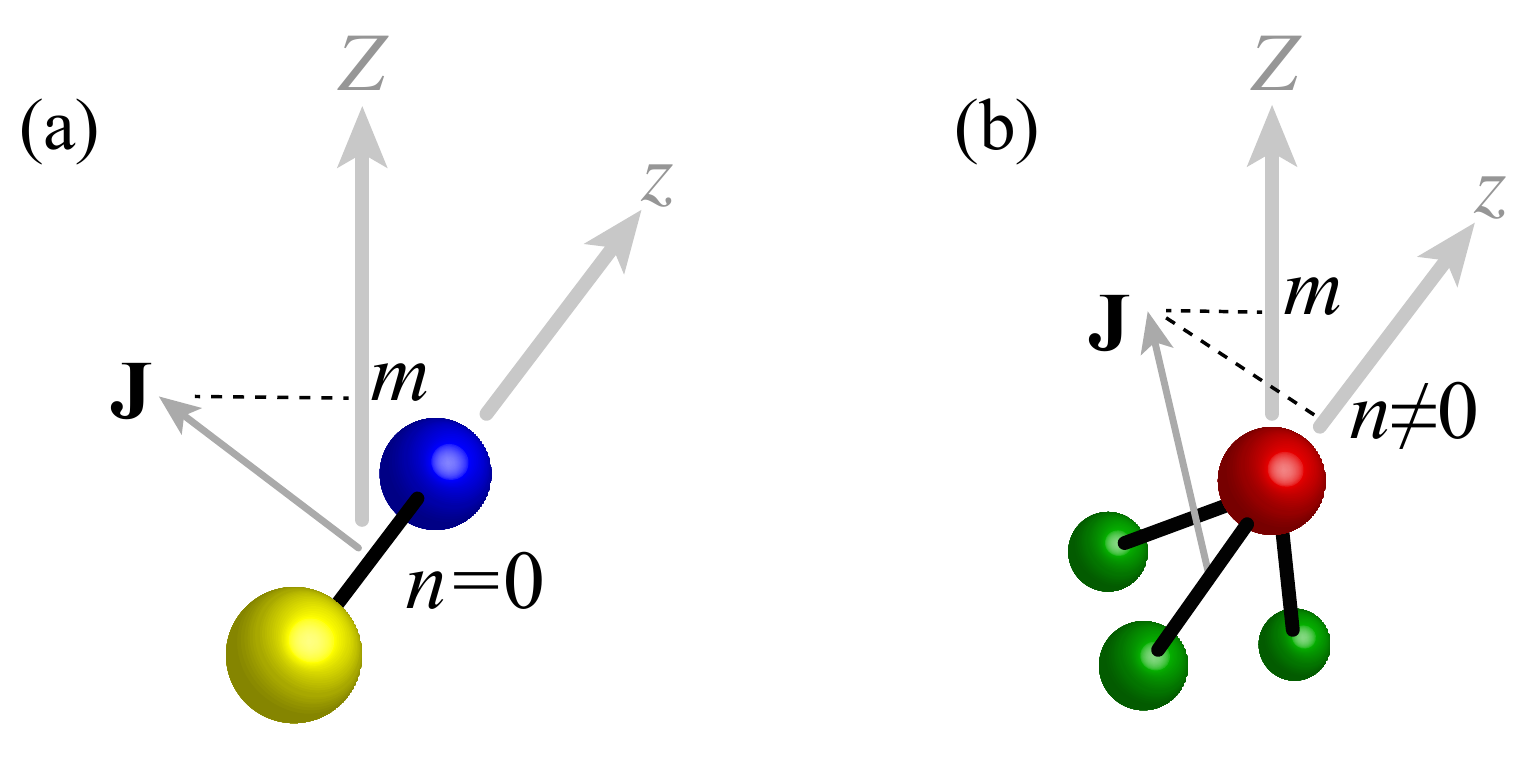}
  \caption{\label{molecules} (a) In a linear-rotor molecule, the rotational angular momentum $\mathbf{J}$ is always perpendicular to the molecular-frame $z$-axis. (b) This is not the case for nonlinear molecules (symmetric and asymmetric tops), which results in a third quantum number, $n$, used to classify the rotational states.}
 \end{figure}

For symmetric-top molecules, the Hamiltonian \eqref{Hrot1} reduces to:
\begin{equation}
\label{Hrot3}
\hat H_\text{rot} =  B \hat J_{x}'^2 + B \hat J_{y}'^2 + C \hat J_{z}'^2 =  B \hat{\mathbf{J}}^2 + G  \hat J_{z}'^2,
\end{equation}
where $G=C-B$. The values of $G<0$ correspond to prolate (cigar-shaped) symmetric tops, while $G>0$ corresponds to the oblate (disk-shaped) ones.

In order to describe the rotation of a symmetric-top molecule, one needs to involve both the laboratory and molecular coordinate systems, and work with both the corresponding angular momentum operators, $ \hat{J}_i$ and  $\hat{J}_i'$. The reason is that, as opposed to linear rotors, the angular momentum vector can have a finite projection on the $z$-axis of the molecular coordinate system, which needs to be taken into account;  see Fig.~\ref{molecules} for a schematic illustration.

Thus, in addition to the quantum numbers we used for a linear rotor, Eq.~\eqref{Jeigen}, we have to introduce an additional quantum number $n$, which gives the projection of $\mathbf{J}$ on the molecular axis:
\begin{align}
\label{JeigenSym}
 \mathbf{\hat{J}^2} \vert j, m, n \rangle &= j(j+1) \vert j, m, n \rangle\\
  \hat{J}_Z \vert j, m, n \rangle &= m \vert j, m, n \rangle \\
    \hat{J}'_z \vert j, m, n \rangle &= n \vert j, m, n \rangle
\end{align}
More details on the properties of the $ \hat{J}_i$ and  $\hat{J}_i'$ operators are provided in Appendix~\ref{sec:appendixAngular}.

In the angular representation, the symmetric-top wavefunctions are given by Wigner $D$-matrices:
\begin{equation}
\label{JDlm}
\langle \phi,  \theta, \gamma \vert j, m,n \rangle = \sqrt{\frac{2j+1}{8 \pi^2} }D^{j \ast}_{m n} (\phi, \theta, \gamma)
\end{equation}
{Note that Eq.~\eqref{JDlm} describes the molecular wavefunction  (i.e.\ it is a function of a set of numbers)  and hence there are no operator hats on top of the angles.} The Wigner $D$-matrices correspond to the rotation operator connecting the space-fixed and body-fixed systems, which are rotated with respect to each other by the Euler angles $(\phi,  \theta, \gamma)$.
A linear rotor corresponds to the limit of a prolate symmetric top whose third moment of inertia $C = 0$. This corresponds to setting $n=0$ in Eqs.~\eqref{JeigenSym} and \eqref{JDlm}, which then reduce to Eqs.~\eqref{Jeigen} and \eqref{JYlm} for the linear rotor. 

\subsubsection{Asymmetric-top molecules}

For asymmetric-top molecules the Hamiltonian \eqref{Hrot1} can be rewritten as~\cite{BernathBook, TownesSchawlow}:
\begin{multline}
\label{HrotAsym}
\hat H_\text{rot} = \left( \frac{A+B}{2} \right) (\hat J_{x}'^2 + \hat J_{y}'^2) + C \hat J_{z}'^2 + \left( \frac{A-B}{2} \right) (\hat J_{x}'^2 - \hat J_{y}'^2) \\
= \left( \frac{A+B}{2} \right) (\hat J_{x}'^2 + \hat J_{y}'^2) + C \hat J_{z}'^2 + \left( \frac{A-B}{2} \right) (\hat J_{-1}'^2 + \hat J_{+1}'^2), 
\end{multline}
{where the ladder operators $\hat J_{\pm 1}'$ are defined by Eqs.~\eqref{Jplusx} and~\eqref{Jminusx}.}

The eigenstates of asymmetric top molecules are written as linear combinations of the symmetric-top states:
\begin{equation}
\label{AsymState}
\vert j, m, i \rangle = \sum_n a_n^i \vert j, m, n \rangle
\end{equation}
The coefficients $a_n^i$ can be obtained by diagonalizing the matrix of $\hat H_\text{rot}$, Eq.~\eqref{HrotAsym}, in the basis of the symmetric-top states, with the matrix elements obtained using Eq.~\eqref{JiPrimeKet} of  Appendix~\ref{sec:appendixAngular}.

\subsection{Boson Hamiltonian}

\subsubsection{Introduction to second quantization}

The most convenient way to work with a many-particle system, such as a BEC, is using the language of second quantization. Here one introduces operators, $\ad_l$ and $\a_l$ which correspond, respectively, to the creation and annihilation of a boson in a state $l$. {Here $l$ is a placeholder for any possible state relevant for a particular problem, and hence can represent an entire set of quantum numbers. For instance, $l$ might simply refer to the lattice site $i$ of an atom, its momentum $\vecp$, its spin state $({\uparrow,\downarrow})$, or angular momentum state $(l,m)$.} Let us consider a many-particle state, where each single-particle state $l$ is occupied by $n_l$ bosons. Then the action of the creation and annihilation operators is defined as follows:
\begin{align}
\a_l \ket{n_1, n_2, \dots, n_l, \dots} &= \sqrt{n_l} \ket{n_1, n_2, \dots, n_l-1, \dots} \\
\ad_l \ket{n_1, n_2, \dots, n_l, \dots} &= \sqrt{n_l+1} \ket{n_1, n_2, \dots, n_l+1, \dots}
\end{align}

 Thus, the action of the annihilation operator on the vacuum state $\ket{0}$ gives zero:
\begin{equation}
\a_l \ket{0} = 0
\end{equation}

One can define a number operator $\n_l = \ad_l \a_l$, with the following property 
\begin{equation}
\n_l \ket{n_1, n_2, \dots, n_l, \dots} =  n_l \ket{n_1, n_2, \dots, n_l, \dots}
\end{equation}

 If $l$ is a discrete variable, such as the position of an atom on a lattice, the commutation relation for the operators is given by:
\begin{equation}
[\a_l, \ad_{l'}]  = \delta_{ll'}
\end{equation}

For free particles, $l$ can be a continuous variable describing the spatial degrees of freedom, such as  a coordinate, $\mathbf{r}$, or momentum, $\mathbf{k}$. The operators in the coordinate and momentum space are related to each other through the Fourier transformation, which corresponds to a change in the basis of the single-particle states, 
\begin{equation}
\label{adFourier}
\ad_\vecr=\int\frac{d^3 k}{(2\pi)^3}\ad_\veck e^{i\veck\vecr}
\end{equation}
\begin{equation}
\label{aFourier}
\a_\vecr=\int\frac{d^3k}{(2\pi)^3}\a_\veck e^{-i\veck\vecr}
\end{equation}

 With $\mathbf{r}$ and $\mathbf{k}$ being continuous variables, the commutation relations are modified according to:
\begin{equation}
\label{ArComm}
	[\a_\mathbf{r}, \ad_\mathbf{r'}] = \delta(\mathbf{r-r'})
\end{equation}
\begin{equation}
\label{ArComm}
	[\a_\mathbf{k}, \ad_\mathbf{k'}] = (2\pi)^3\delta(\mathbf{k-k'})
\end{equation}
We note that the prefactor of $(2\pi)^3$ is  a matter of  convention we adopt here, and varies across literature. {The convention used here allows to avoid unnecessary prefactors in the commutators of the operators in the angular momentum representation, see Eq.~\eqref{AklmComm} below.}
 
 \subsubsection{A system of interacting bosons}

 Let us now consider a system of interacting bosons without a molecule being present. Its   Hamiltonian can be written in momentum representation as:
\begin{equation}
\label{Hbos}
\hat H_\text{bos}= \sum_\mathbf{k}  \epsilon (\veck) \ad_\mathbf{k} \a_\mathbf{k} +\frac{1}{2} \sum_\mathbf{k, k', q} V_\text{bb} (\mathbf{q})\ad_\mathbf{k'-q} \ad_\mathbf{k+q} \a_\mathbf{k'}  \a_\mathbf{k}
\end{equation}
{Here we introduced the label $\sum_\mathbf{k}  \equiv \int d^3k/(2\pi)^3$ in order to keep the notation compact. Since we work in units where $\hbar\equiv1$, momentum carries units of wavenumbers, [Length]$^{-1}$, and each summation carries a dimensionality of [Length]$^{-3}$}. The first term of Eq.~\eqref{Hbos} describes  the kinetic energy of bosons, $\epsilon (\veck) = k^2/(2 m)$, with $m$ the bosonic mass. The second term gives the boson-boson interactions, whose strength in momentum space is given by $ V_\text{bb} (\mathbf{q})$. Note that  in our convention the operators $\ad_\mathbf{k}$ and $\a_\mathbf{k}$ are not dimensionless, but carry a dimension of [Length]$^{3/2}$. {Furthermore, since the Fourier transform involves three-dimensional integration in real space, the momentum-dependent interaction potential $V_\text{bb} (\mathbf{q})$ carries a unit of [Energy]$\times$[Length]$^{3}$}.

 The Hamiltonian~\eqref{Hbos} represents the most general  formulation of the problem. Due to its complexity, it cannot be solved analytically and one has to resort to approximate methods.  In what follows we discuss the possible approximations  which allow to obtain insight into the behavior of such a complex many-particle system.

\subsubsection{Bogoliubov approximation and transformation}
 \label{sec:bogoliubov}
 
Let us assume a weakly-interacting BEC at zero temperature. In such a case, most of the atoms reside in  their energetically lowest single particle state {\cite{Pitaevskii2016}}, which for an infinite system is the zero-momentum quantum state.  In order to make use of this separation of occupation numbers it is convenient to split the creation and annihilation operators into the zero-momentum and finite-momentum parts:
\begin{equation}
\label{BogApp}
\a_\mathbf{k} = (2\pi)^3 \hat \Phi_0 \delta(\mathbf{k}) + \B_\mathbf{k \neq 0}
\end{equation}
{where the factor of $(2\pi)^3$ appears due to our definition of the Fourier transform, Eqs.~\eqref{adFourier},~\eqref{aFourier}.}
 This formally exact shift  of the creation and annihilation operators (which relies on the description of the BEC state as a coherent state) allows to treat in a convenient way the excitations on top of this state as  small perturbations. Within the Bogoliubov approximation~\cite{Pitaevskii2016}, one assumes that the fraction of the bosons in the condensate, $\hat \Phi_0$, is very large, and corresponds to the classical number density of particles $n$.\footnote{Strictly speaking, $\hat \Phi_0$ corresponds to the BEC density $n_0$ which, however, for most practical purposes can be replaced by $n$ \cite{Pitaevskii2016}.} As a consequence, the field operators corresponding to the $\mathbf{k}=0$ state are replaced as  $\hat \Phi_0 \to \sqrt{n}$.

The second part of the approximation makes use of the fact that the population of excited states is small, $\sum_\mathbf{k} \langle |\B_\mathbf{k}|^2 \rangle \ll n$. As a result, upon substitution of Eq.~\eqref{BogApp} into Eq.~\eqref{Hbos}, one can neglect the terms cubic and quartic in  $\B_\mathbf{k}$. Then, assuming $V_\text{b-b}(\mathbf{q}) = V_\text{b-b}(-\mathbf{q})$,  and dropping the constant terms, we obtain: 
 \begin{equation}
\label{HbigB}
	\hat H_\text{bos}=  \sum_\mathbf{k}  \left[ \epsilon (\veck) + V_\text{bb} (\mathbf{k}) n \right] \Bd_\mathbf{k} \B_\mathbf{k} +  \frac{n}{2} \sum_\mathbf{k}   V_\text{b-b}(\mathbf{k}) \left [ \Bd_\mathbf{k} \Bd_\mathbf{-k} +  \B_\mathbf{k} \B_\mathbf{-k}    \right] \end{equation}
	
Thus, we have arrived at a quadratic Hamiltonian, Eq.~\eqref{HbigB}{. In the second term this Hamiltonian contains, however, anomalous terms where two creation or two annihilation   operators  are paired together. To eliminate these terms,  a  diagonalization in the space of creation and annihilation operators can be performed, which at the same time diagonalizes the Hamiltonian in momentum space}. This is achieved by means of the so-called Bogoliubov transformation -- also called `Bogoliubov rotation' --  of the field operators~\cite{Pitaevskii2016}:
\begin{align}
\label{BogolRot}
	\B_\mathbf{k} &= u_\veck \be_\mathbf{k} + v^\ast_{-\veck} \bed_\mathbf{-k}\\
	\Bd_\mathbf{k} &= u^\ast_\veck \bed_\mathbf{k} + v_{-\veck} \be_\mathbf{-k}
\end{align}
 where the coefficients $u_\veck$ and $v_\veck$ obey the  normalization condition:
\begin{equation}
\label{BogNorm}
	|u_\veck|^2 - |v_\veck|^2 = 1
\end{equation}

For simplicity, we assume $u_\veck$, $v_\veck$ to be real, symmetric with respect to $\veck \to -\veck$, and dependent solely on $k=|\veck|$ leading to the following expressions that diagonalize the Hamiltonian {(note that this does not hold e.g.\ if the Bose gas is unstable, confined in a harmonic trap, or in the presence of vortices)}:
\begin{align}
\label{ukvkomega}
	u_k  &= \left( \frac{\epsilon (\veck) + V_\text{bb}(\mathbf{k}) n}{2 \omega (\veck)} + \frac{1}{2} \right)^{1/2} \\
	v_k  &= - \left( \frac{\epsilon (\veck) + V_\text{bb}(\mathbf{k}) n}{2 \omega (\veck)} - \frac{1}{2} \right)^{1/2}
\end{align}
{where $\omega (\veck) \equiv \omega (k)$ is given by:
\begin{equation}
\label{Bogwk}
	\omega (k) = \sqrt{\epsilon (k) \left[ \epsilon (k) + 2 V_\text{bb}(k) n \right]}
\end{equation}
}
After the transformation, the boson Hamiltonian, Eq.~\eqref{HbigB}, takes the diagonal form:
\begin{equation}
\label{Hwk}
	\hat H_\text{bos}=  \sum_{\veck} \omega (k) \bed_\mathbf{k}  \be_\mathbf{k}
\end{equation}
 Thus, the Bogoliubov transformation allows to describe the bosonic systems in terms of non-interacting Bogoliubov quasiparticles with a dispersion relation {$\omega (k)$.}
 
 For a dilute atomic BEC, the boson-boson interaction can be approximated well by a contact potential, so that in first order Born approximation {one can replace $V_\text{bb} (k)$ by a constant, $g_\text{bb}=4\pi a_{\text{bb}}/m$,} where  $a_{\text{bb}}$ is the boson-boson scattering length~\cite{Pitaevskii2016}.  In order to avoid the collapse of the BEC, we assume weak effective repulsive interactions characterized by $a_{\text{bb}}>0${, such that they do not significantly disturb the BEC state, yet do prevent it from a collapse}.  The dispersion relation $\omega (k)$ for such a situation is schematically shown in Fig.~\ref{dispersion}(a) by the blue line. Depending on the quasiparticle momentum $k$, one can clearly distinguish two types of behavior. At small momenta $k$, the behavior is linear,  $\omega (k) \approx c k$, where $c = \sqrt{g_\text{bb} n /m}$ is the speed of sound. This regime corresponds {to collective long-wavelength excitations of the condensate, which are called `phonons.'} At large momenta, i.e.\ the ones exceeding the inverse  healing length of the BEC, $\xi^{-1} = \sqrt{2} m c /\hbar$, the dispersion coincides with the one for free particles, $\omega (k) \approx \epsilon (k)$, and one speaks of `particle-like' excitations. {However, in what follows, we will refer to both particle-like and wave-like excitations described by Eq.~\eqref{Bogwk} as `phonons.'}
 
Compared to ultracold gases, superfluid helium is extremely dense. {As a consequence, the average interparticle distance between the atoms in helium, $\langle r \rangle \sim 4$~\AA, is small enough to lie within the short-range part of the He--He potential energy surface, which possesses a minimum at $r_m \approx 3$~\AA. Therefore, He--He interactions cannot be well described as being point-like in real space when compared to the average interatomic distance.} As a result, the dispersion relation changes qualitatively, as one can see by comparing the blue and red lines in Fig.~\ref{dispersion}(a). The most drastic difference is the appearance of a `roton minimum'  which emerges for helium around $k=2.0$~\AA$^{-1}$. It is important to note that, unlike first conjectured by Landau and Feynman, rotons do not represent elementary quanta of rotational excitations, or vortices.  Instead, rotons  can be thought of as density-wave excitations with a finite wave vector,  in this sense they represent precursors or  `ghosts' of crystalline structure. Furthermore, rotons are not specific to superfluid helium, in fact, they arise  for any interactions $V_\text{bb}(k)$ which change the sign as a function of $k$. Such interactions appear, for instance, between  Rydberg-dressed atoms~\cite{Henkel2010, OtterLemPRL14} as well as  in ultracold dipolar~\cite{LahayePfauRPP2009} and quadrupolar~\cite{LarzNJP15} gases confined in quasi-1D and quasi-2D geometries.  In order to illustrate how the roton minimum emerges from Eq.~\eqref{Bogwk},  let us consider the simplest possible toy model featuring such sign-changing interactions.  Let us assume a boson-boson interaction described, for instance,  by the Bessel function of the first kind, $V_\text{bb}(k)  = J_0 (k)$ {(alternatively, one could choose any other sign-changing function with a magnitude decaying at $k \to \infty$):}
\begin{equation}
\label{wkModel}
		\omega_\text{model} (k) = \sqrt{k^2 \left[k^2 + 2 J_0 (k) n \right]}
\end{equation}
Since here we are interested solely in the qualitative properties of the dispersion relation, we omit all the constant factors such as the boson mass $m$. As a result, the momentum $k$ and the density $n$ of Eq.~\eqref{wkModel} are expressed in some arbitrary units.  Such a sign-changing model interaction $V_\text{bb}(k)$  is shown in Fig.~\ref{dispersion}(b) by the red dashed line. The other lines correspond to $\omega_\text{model} (k)$ calculated at different values of density $n$.  At lower densities, such as $n=1$ (in the arbitrary units used here), the sign changing term is negligible, $2 |J_0 (k)| n \ll k^2$, and therefore the dispersion relation looks similar to the one for a weakly-interacting BEC. When $n$ increases to $10$ and $15$, however, the sign-changing dependence of the potential starts to play a crucial role and the roton minimum arises in the dispersion relation. As the density is increased further, the minimum of the dispersion relation eventually touches zero. From this point on the dispersion relation becomes imaginary {for a finite range of momenta.} This situation reveals  an instability  of the system where the rotons materialize and form a  crystalline structure. Note that here $n$ plays a role of a mathematical parameter (expressed in arbitrary units), which regulates the effective strength of the interatomic interactions. Therefore, the values considered in Fig.~\ref{dispersion}(b) do not directly correspond to any particular experimental scenario. However, in a qualitatively similar way, such a transition leads to crystallization of liquid helium if the density $n$ is increased by applying external pressure~\cite{AtkinsHelium}.

We note that for finite systems, such as helium droplets, other types of excitations can  appear. One  example are ripplons -- quantized surface waves or `wrinkles' on the droplet surface~\cite{HansenPRB07, StienkemeierJPB06}. However, whether one works with molecules in small droplets, bulk helium, or a dilute BEC, many of the properties of the collective excitations are contained in the dispersion relation $\omega (k)$. Therefore, quite often, the theory can be constructed assuming a general form of the dispersion relation, which allows to understand the similarities and distinctions between different systems.

\begin{figure}[t]
  \centering
 \includegraphics[width=0.8\linewidth]{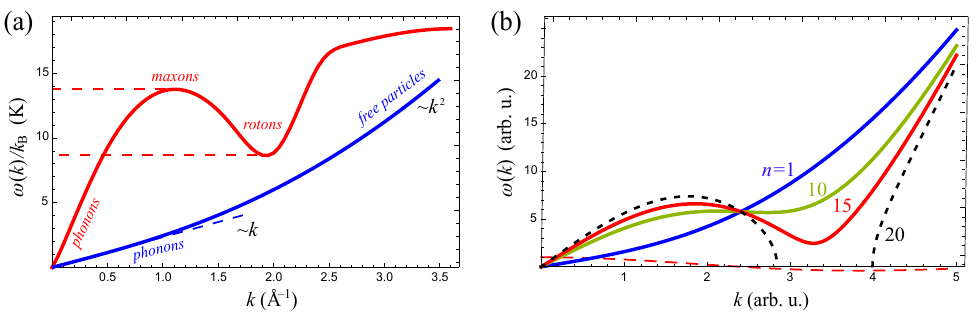}
  \caption{\label{dispersion} (a) Experimental dispersion relation {$\omega(k)$} of superfluid $^4$He, from Ref.~\cite{DonnellyHe98} (red line), compared to a schematic dispersion relation {$\omega(k)$} for a  weakly-interacting BEC with contact interactions, Eq.~\eqref{Bogwk} with $ V_\text{bb}(k) = g_\text{bb}$~{(blue line)}. {Different types of excitations taking place are indicated: phonons, maxons, rotons, and free particles.} (b) Emergence of the roton minimum {(solid blue to solid red)} and instability {(dashed black)} for the model dispersion relation, Eq.~\eqref{wkModel}{, for different values of the density parameter $n$.} The red dashed line shows the sign-changing model potential  $ V_\text{bb}(k) = J_0 (k)$.}
 \end{figure}

  \subsubsection{Angular-momentum representation of the boson operators}
  
In the previous sections, we expressed the boson creation and annihilation operators, $\bed_\veck$ and $\be_\veck$, in Cartesian coordinates, $\veck = \{k_x, k_y, k_z\}$, which may be rewritten in the spherical basis of $\veck = \{k, \Theta_k,\Phi_k\}$. However, we are interested in rotating impurities and the angular momentum properties of the condensate. For that reason, it is much more convenient to work in the angular momentum representation  for the single-particle basis instead of the Cartesian one. Hence we perform a single-particle basis change which yields the following transformation of the creation operators:
\begin{equation}
\label{AklmAk}
	\bd_{k\lambda \mu} =\frac{k}{(2\pi)^{3/2}} \int   d\Phi_k d\Theta_k~\sin\Theta_k~\bd_\mathbf{k} ~ i^{-\lambda}~ Y_{\lambda \mu} (\Theta_k, \Phi_k)
\end{equation}
\begin{equation}
\label{AkAklm}
	\bd_\mathbf{k} = \frac{(2\pi)^{3/2}}{k} \sum_{\lambda \mu}  \bd_{k\lambda \mu}~ i^{\lambda}~Y^\ast_{\lambda \mu} (\Theta_k, \Phi_k), 
\end{equation}
 Here, the  quantum numbers $\lambda$ and $\mu$ label the angular momentum of the bosonic excitation and its projection onto the laboratory-frame $Z$-axis, respectively.  Note that the convention of~Eqs.~\eqref{AklmAk}, \eqref{AkAklm} differs from the one used in Refs.~\cite{SchmidtLem15, SchmidtLem16} by the conjugates of the spherical harmonics and the phase factor. The convention we use in this tutorial is thereby made consistent with other sources, such as Ref.~\cite{GreinerBook}. Within this convention, the creation and annihilation operators fulfill the  following commutation relations:
\begin{equation}
\label{AkComm}
	[\be_\mathbf{k}, \bd_\mathbf{k'}] = (2\pi)^3\delta^{(3)}(\mathbf{k-k'})
\end{equation}
\begin{equation}
\label{AklmComm}
	[\be_{k\lambda \mu}, \bd_{k'\lambda' \mu'}] = \delta(k-k') \delta_{\lambda \lambda'} \delta_{\mu \mu'}
\end{equation}
 Within the same convention as for the momentum representation, in coordinate space, the Cartesian and angular momentum representations are related by:
\begin{equation}
\label{ArlmAr}
	\bd_{r\lambda \mu} = r  \int d\Phi_r d\Theta_r~\sin\Theta_r ~\bd_\mathbf{r}~Y_{\lambda \mu} (\Theta_r,\Phi_r)
\end{equation}
\begin{equation}
\label{ArArlm}
	\bd_\mathbf{r} = \frac{1}{r} \sum_{\lambda \mu}  \bd_{r\lambda \mu} ~Y_{\lambda \mu}^\ast (\Theta_r,\Phi_r)
\end{equation}
with the corresponding commutation relations:
\begin{equation}
\label{ArComm}
	[\be_\mathbf{r}, \bd_\mathbf{r'}] = \delta^{(3)}(\mathbf{r-r'})
\end{equation}
\begin{equation}
\label{ArlmComm}
	[\be_{r \lambda \mu}, \bd_{r' \lambda' \mu'}] = \delta(r-r') \delta_{\lambda  \lambda'} \delta_{\mu \mu'}
\end{equation}
Since the Cartesian operators in  the coordinate and momentum space are related through the Fourier transformation,
\begin{equation}
\label{brViabk}
	\be^\dagger_\mathbf{r} = \int \frac{d^3  k}{(2\pi)^3} \be^\dagger_\mathbf{k}  e^{-i \mathbf{k \cdot r}},
\end{equation}
one can obtain the corresponding relation for their angular momentum components:
\begin{equation}
\label{brlmViabklm}
	\be^\dagger_{r \lambda \mu} = \sqrt{\frac{2}{\pi}} r \int k dk~j_\lambda (kr)~\be^\dagger_{k\lambda\mu},
\end{equation}
 where  $j_\lambda (kr)$ is the spherical Bessel function~\cite{AbramowitzStegun}.

 \subsection{Molecule-boson interaction}

In its most general form, the interaction between an impurity and the  bosonic atoms is given by:
 \begin{equation}
\label{Himpbos}
 \hat H_\text{mol-bos}  =\sum_\mathbf{k, q}  \hat V_\text{mol-bos} (\mathbf{q}, \hat{\phi},  \hat{\theta}, \hat{\gamma})   \hat \rho(\mathbf{q}) \ad_\mathbf{k+q} \a_\mathbf{k},
\end{equation}
 where $\hat \rho(\mathbf{q}) = e^{- i \mathbf{q} \hat{\mathbf{r}}}$  is the Fourier-transformed density of an impurity which is situated at position $\hat \vecr$ (the corresponding density in real space is given by a Dirac $\delta$-function). Here we consider an impurity whose translational motion is frozen, and which we position at the coordinate $\vecr=0$ (for an impurity translationally moving in space the factor of  $\hat\rho(\mathbf{q}) = e^{- i \mathbf{q}\hat{\mathbf{r}}}$ results in a hybrid problem between the angulon and the Fr\"ohlich polaron~\cite{MidyaInPrep}).  This choice results in $\hat\rho(\mathbf{q}) \equiv 1$. The molecule has, however, additional degrees of freedom corresponding to the orientation of its  molecular axis, and, depending on the molecular orientation, the surrounding bosons see a different interaction potential.   The fact that anisotropic molecular geometry gives rise to anisotropic molecule-boson interactions is represented in Eq.~\eqref{Himpbos} by the operator $\hat V_\text{mol-bos} (\mathbf{q}, \hat{\phi},  \hat{\theta}, \hat{\gamma}) $ which explicitly depends on the orientation of the molecule in space, as given by the Euler angle operators $(\hat{\phi},  \hat{\theta}, \hat{\gamma})$.  It is important to note that as opposed to the polaron problem~\cite{Devreese13}, the interactions considered here are not spherically symmetric, therefore the replacement of $\mathbf{q} \to \mathbf{-q}$ in Eq.~\eqref{Himpbos} results in an additional phase factor in the final expression of Eq.~\eqref{HintFinal}.

Let us now derive an explicit expression for  $ \hat V_\text{mol-bos}(\mathbf{q}, \hat{\theta}, \hat{\phi}) $, which is the Fourier transform of the molecule-boson potential in real space, starting from first principles. For simplicity, here we assume a linear rotor molecule, whose orientation is given by only two angles, $(\hat{\theta}, \hat{\phi})$. For symmetric and asymmetric tops the derivations will have a similar form, however, with a dependence on the third Euler angle, $\hat \gamma$.
As previously, we define two coordinate frames: the laboratory  frame, $(X,Y,Z)$,  in which the bosonic atoms are at rest when the molecule is absent, and the molecular one, $(x, y, z)$, used to define the microscopic molecule-boson  interaction potential, see Fig.~\ref{rotor}(a).

The interaction potential between a molecule and an atom is a function of the spherical coordinates in the molecular frame, $(\theta_r, \phi_r)${, as schematically shown in Fig.~\ref{rotor}(b). Such a potential  can be  expanded over the spherical harmonics as:}
\begin{equation}
\label{Vr}
	V_\text{mol-bos} (\mathbf{r}) = \sum_\lambda  V_\lambda (r) Y_{\lambda 0} (\theta_r, \phi_r),
\end{equation}
{Here, the interaction in every angular momentum channel $\lambda$  is defined by $V_\lambda (r)$.} Usually, the interaction potentials~\eqref{Vr} are obtained using quantum chemistry computer codes~\cite{SzalewiczIRPC08}, or approximated by some analytic model functions. 

In order to transform Eq.~(\ref{Vr}) to the laboratory frame, where the bosonic part of the Hamiltonian is defined, we use Wigner rotation matrices~\cite{VarshalovichAngMom, ZareAngMom}:
\begin{equation}
\label{WignerD}
	Y_{\lambda 0} (\theta_r, \phi_r) = \sum_\mu  \hat D^{\lambda}_{\mu 0} (\hat{\phi}, \hat{\theta}, \hat{\gamma}) Y_{\lambda \mu} (\Theta_R, \Phi_R) 
\end{equation}
 For a linear molecule the third angle, $\hat{\gamma}$, can be set to zero. In this case, given that $\hat D^{\lambda}_{\mu 0} (\hat{\phi}, \hat{\theta}, 0) = \sqrt{\frac{ 4 \pi}{2 \lambda +1}} \hat Y_{\lambda \mu}^\ast (\hat{\theta}, \hat{\phi})$~\cite{VarshalovichAngMom, ZareAngMom}, we obtain:
\begin{equation}
\label{Vr2}
	\hat V_\text{mol-bos}(\mathbf{R}, \hat \theta,\hat \phi) = \sum_{\lambda \mu} \sqrt{\frac{ 4 \pi}{2 \lambda +1}}  V_\lambda (r)  Y_{\lambda \mu} (\Theta_R, \Phi_R)  \hat Y_{\lambda \mu}^\ast (\hat \theta,\hat \phi)
\end{equation}
Here $\mathbf{R} \equiv (R, \Theta_R, \Phi_R)$ gives the laboratory-frame coordinates of a boson interacting with the molecule{, whose axis' orientation is measured by the operators $(\hat \theta,\hat \phi)$}.

{It is important to note that the procedure described above is quite general and is often used to describe molecular rotation in the presence of an electromagnetic field acting in the laboratory frame. Here, however, we deal with  a many-particle field of bosonic atoms, which makes the problem more involved.}

\begin{figure}[t]
  \centering
 \includegraphics[width=0.5\linewidth]{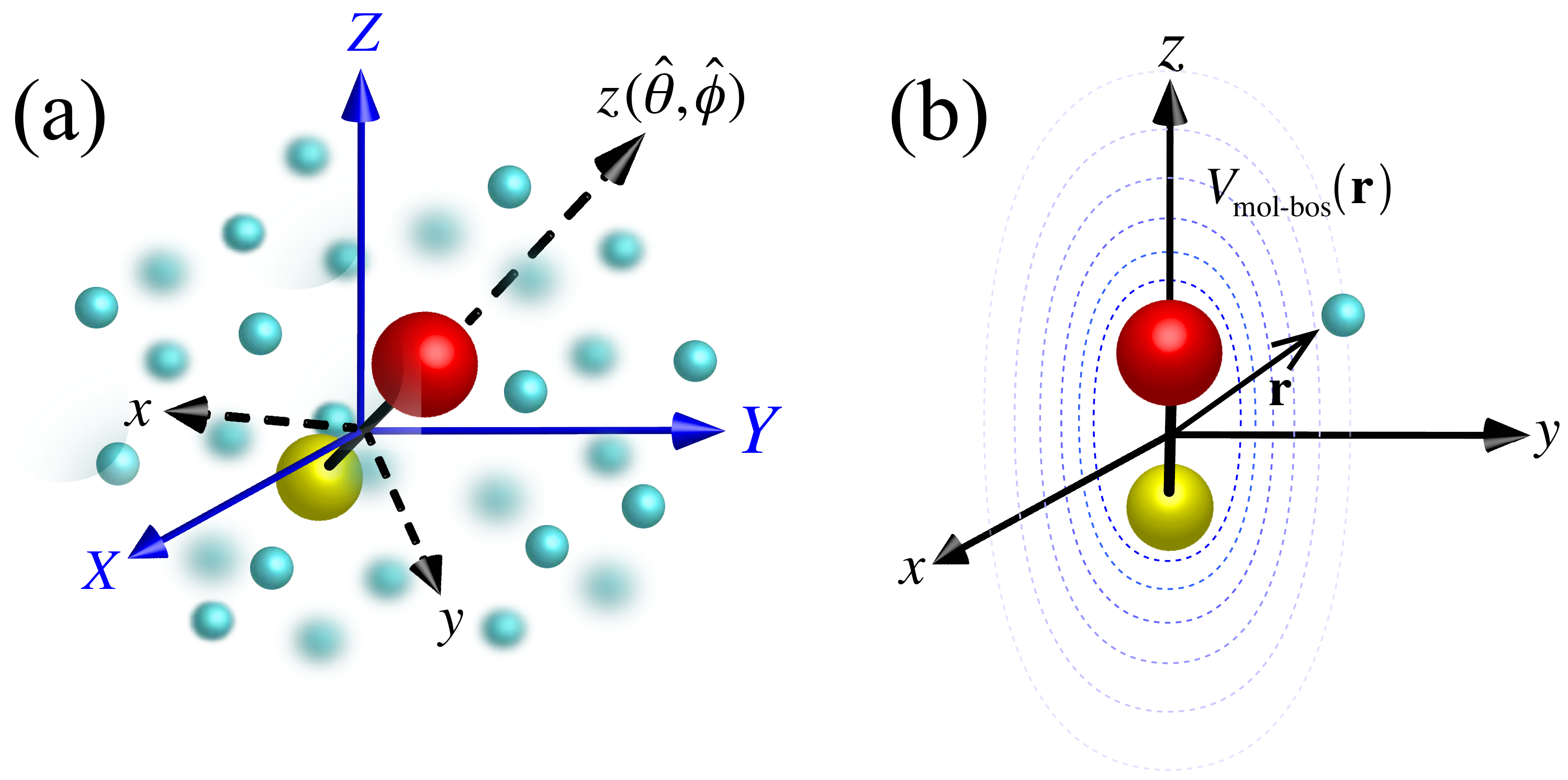}
  \caption{\label{rotor} (a) A linear-rotor molecule immersed into a boson bath. The molecule-boson interaction explicitly depends on the rotor angular coordinates, $(\hat{\theta}, \hat{\phi})$, in the laboratory frame. (b) The anisotropic molecule-boson interaction is defined in the molecular coordinate frame. Adapted with permission from Ref.~\cite{SchmidtLem15}.}
 \end{figure}
 
The interaction in the momentum space is given by the Fourier transform of Eq.~(\ref{Vr2}):
\begin{equation}
\label{Vk}
	\hat V_\text{mol-bos}(\mathbf{k}, \hat \theta,\hat \phi) \equiv \int d^3R~\hat V_\text{mol-bos}(\mathbf{R}, \hat \theta,\hat \phi) e^{-i \mathbf{k R}} =  \sum_{\lambda \mu} (2\pi)^{3/2}  i^{-\lambda} \tilde{V}_\lambda (k)  Y_{\lambda \mu} (\Theta_k, \Phi_k)  \hat Y_{\lambda \mu}^\ast (\hat{\theta}, \hat{\phi}),
\end{equation}
where $\mathbf{k}  \equiv (k, \Theta_k, \Phi_k)$ is the momentum vector in the laboratory frame and
\begin{equation}
\label{Vtilde}
\tilde{V}_\lambda (k) =   2^{3/2}/\sqrt{2\lambda + 1} \int_0^\infty  dr\, r^2 V_\lambda(r) j_\lambda (kr)
\end{equation}
In order to derive Eq.~\eqref{Vk}, we made use of the plane wave expansion in spherical harmonics:
\begin{equation}
\label{PlaneWave}
	e^{-i\veck \mathbf{R}} =\sum_{lm}4 \pi~i^{-l}~j_l(kR)Y^\ast_{lm}(\Theta_R, \Phi_R)Y_{lm}(\Theta_k, \Phi_k)
\end{equation}
 
As a next step, we apply the Bogoliubov approximation and transformation {(see Sec.~\ref{sec:bogoliubov})} to Eq.~\eqref{Himpbos} and obtain:
\begin{multline}
\label{BogImp}
\hat H_\text{mol-bos} = n  \hat V_\text{mol-bos} (\mathbf{k}= 0, \hat \theta,\hat \phi) + \sqrt{n} \sum_\veck \hat V_\text{mol-bos}(\mathbf{k}, \hat \theta,\hat \phi)   (\Bd_\mathbf{k}+\B_\mathbf{-k}) \\
 	= n  \hat V_\text{mol-bos} (\mathbf{k}= 0, \hat \theta,\hat \phi) + \sqrt{n} \sum_\veck \hat V_\text{mol-bos}(\mathbf{k}, \hat \theta,\hat \phi) \sqrt{\frac{\epsilon (\veck)}{\omega (\veck) }}  (\bed_\mathbf{k} + \be_\mathbf{-k}).
\end{multline}
Here $\hat V_\text{mol-bos} (\mathbf{k}= 0, \hat \theta,\hat \phi)$ represents a mean-field shift, whose magnitude does not depend on the molecular orientation in space. Such a constant, spherically symmetric contribution is the same as  for linearly moving polarons, and provides equal shift to all angular momentum levels of the molecule. Therefore, hereafter it will be omitted.
It is important to note that   in  Eq.~\eqref{BogImp}, we neglected  terms quadratic in the bosonic excitations at finite momentum. While in some situations these terms are important~\cite{Rath2013, Shchadilova2016, Jorgensen2016,Ashida2017}, in the limit of a large number of particles, $n \to \infty$, and in the absence of resonant impurity-bath interactions, the linear term will dominate~\cite{GirardeauPF61, llp}.

As a final step, we substitute into Eq.~\eqref{BogImp} the spherical representation of the boson operators, Eq.~\eqref{AklmAk}, as well as the explicit expression for the interaction potential in momentum space, Eq.~\eqref{Vk}. After integrating over angles (and using the orthogonality relations for spherical harmonics~\cite{VarshalovichAngMom}), we obtain:
 \begin{equation}
\label{HintFinal}
\hat H_\text{mol-bos} = \sum_{k \lambda \mu} U_\lambda(k)  \left[ \hat Y^\ast_{\lambda \mu} (\hat \theta,\hat \phi) \bed_{k \lambda \mu}+ \hat Y_{\lambda \mu} (\hat \theta,\hat \phi) \be_{k \lambda \mu} \right],
\end{equation}
where we labeled $\sum_k\equiv\int dk$ (note the difference with three-dimensional sums of Eq.~\eqref{Hbos}) and 
 \begin{equation}
\label{Ulamk}
U_\lambda(k) =  \left[\frac{8 n k^2\epsilon (k)}{\omega (k)(2\lambda+1)}\right]^{1/2} \int dr r^2  V_\lambda(r) j_\lambda (kr)
\end{equation}

The interaction Hamiltonian \eqref{HintFinal} plays a key role in this tutorial. It is worth noting that  Hamiltonians featuring  a linear coupling between an impurity's internal degree of freedom and a bath of harmonic oscillators have been actively studied, see e.g.\ the spin-boson~\cite{LeggettRMP87} or Jaynes-Cummings~\cite{JanyensCummings} models. In contrast to these models, however, $H_\text{mol-bos}$ explicitly depends on the molecular angle operators, $(\hat \theta,\hat \phi)$, which is essential {for the microscopic description of a rotating anisotropic impurity} and the emergence of angulon physics. {On the other hand, unlike in the Fr\"ohlich Hamiltonian for a linearly moving impurity~\cite{Devreese13}, the spherical harmonic operators $\hat Y_{\lambda \mu} (\hat \theta,\hat \phi)$ of Eq.~\eqref{HintFinal} lead to additional complexity due to the   angular momentum algebra involved.}

{By analogy with Eqs.~\eqref{AngleOp1} and \eqref{DlkiAngle}, we obtain:
 \begin{equation}
\label{YlmAngle}
\hat Y_{\lambda \mu} (\hat \theta,\hat \phi) \ket{\theta, \phi} = Y_{\lambda \mu}(\theta, \phi) \ket{\theta, \phi} 
\end{equation}
Thus we can rewrite the $\hat Y_{\lambda \mu} (\hat \theta,\hat \phi)$ operators in the angular momentum basis, by inserting complete sets in angles, $\hat 1 \equiv \int \sin \theta d\theta d\phi \ket{\theta \phi}\bra{\theta \phi}$, as well as  a complete set in angular momenta, $\hat 1 \equiv \sum_{jm} \ket{jm}\bra{jm}$. After integration over angles one finds~\cite{VarshalovichAngMom}:}
  \begin{equation}
\label{YlmOper}
\hat Y_{\lambda \mu} (\hat \theta,\hat \phi) = \sum_{jmj'm'} a_{jm, \lambda \mu}^{j'm'} \ket{j' m'}\bra{jm}
\end{equation} 
where
  \begin{equation}
\label{YlmOperCoef}
a_{jm, \lambda \mu}^{j'm'} = \sqrt{\frac{(2j+1)(2\lambda+1)}{(2j'+1) 4\pi}} C_{jm, \lambda \mu}^{j'm'} C_{j0, \lambda 0}^{j' 0}
\end{equation} 
Here $C_{jm, \lambda \mu}^{j'm'}$ are the Clebsch-Gordan coefficients~\cite{VarshalovichAngMom}.

From \eqref{YlmOper} one can see that the $\hat Y_{\lambda \mu} (\hat \theta,\hat \phi)$ operators, in principle, couple all rotational states with one another, with the selection rules given by the Clebsch-Gordan coefficients of Eq.~\eqref{YlmOperCoef}.  However, the situation becomes often manageable due to the fact that for realistic potentials in the expansion of the interaction potential, Eq.~\eqref{Vr2}, only a few terms have significant contributions.
For example, interactions of a strongly dipolar molecule with an atom will be dominated by the term with $\lambda=1$. This interaction is analogous to an electric field and will result in the selection rules of $j' = j \pm 1$. For a homonuclear molecule, the quadrupole contribution, $\lambda=2$, will dominate, and consequently $j' = j \pm 2$.

 \section{The angulon quasiparticle}
 \label{sec:angulon}
 
{In this section we combine the ideas outlined above to  finally study the physics of the full Hamiltonian describing a molecular rotor impurity immersed inside a BEC.} For simplicity, we consider the case of a linear-rotor molecule. The resulting equations can, however, be generalized to more complex molecular species. Combining Eqs.~\eqref{Hrot2}, \eqref{Hwk}, and~\eqref{HintFinal} we obtain the following Hamiltonian:
 \begin{equation}
\label{Hamil1}
 \hat H= B \mathbf{\hat{J}^2} + \sum_{k \lambda \mu}  \omega (k) \bed_{k\lambda \mu} \be_{k\lambda \mu} +   \sum_{k \lambda \mu} U_\lambda(k)  \left[ \hat Y^\ast_{\lambda \mu} (\hat \theta,\hat \phi) \bed_{k \lambda \mu}+ \hat Y_{\lambda \mu} (\hat \theta,\hat \phi) \be_{k \lambda \mu} \right]
\end{equation}

It is important to note that although we have initially derived  the specific form of  $\omega (k)$ and $U_\lambda(k)$ in  Eq.~\eqref{Hamil1} for the case of an ultracold molecule coupled to a weakly-interacting BEC, this Hamiltonian can be used to study the transfer of angular momentum between a localized impurity and a bath of harmonic oscillators in the context of other experiments. For instance, effective Hamiltonians of a similar structure can be constructed for molecules in helium droplets~\cite{LemeshkoDroplets16, YuliaPhysics17, Shepperson16}, Rydberg atoms in a BEC~\cite{BalewskiNature13, SchmidtDem2016}, electronic excitations coupled to phonons in solids~\cite{StammPRB10, TsatsoulisPRB16, FahnleJSNM17}, and several other systems. Thus, in what follows, we approach Eq.~\eqref{Hamil1} from a completely general perspective and reveal its properties in various parameter regimes.

\subsection{Second-order perturbation theory}
\label{sec:2dOrder}

{As a first step, we account for the effect of the bath on molecular rotation within second-order perturbation theory, which applies when the interactions between the impurity and the bath are weak,  $|U_\lambda(k)| \ll B$}. There,  one accounts only for  a single virtual phonon excitation from the vacuum $\ket{0}$  to the states {$\vert k \lambda \mu \rangle \equiv  \bed_{k \lambda \mu} \ket{0}$}, accompanied by the change of the molecular rotational state from $\vert j, m\rangle$ to $\vert j', m'\rangle$, and back.  Accordingly, the second-order energy shift acquired by the  state $\vert j, m\rangle$ is given by:

\begin{figure}[b]
  \centering
    \includegraphics[width=0.3\linewidth]{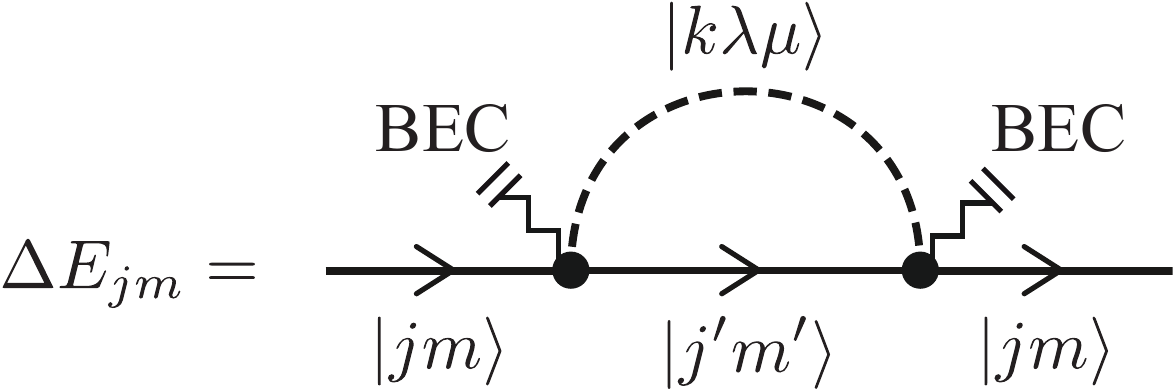}\caption{Feynman diagram representing the second-order perturbation corrections to the angulon energy.}
  \label{sunset}
\end{figure}

\begin{equation}
\label{DEJM}
	\Delta E_{jm} = \sum_{\substack{j'm' \\ k\lambda \mu}} \frac{\Bigl | \bra{k \lambda \mu} \bra{j'm'} H_\text{mol-bos} \ket{jm} \ket{0}  \Bigr |^2}{B j(j+1) - Bj'(j'+1) - \omega (k) } = \sum_{k \lambda j' } \frac{2\lambda +1}{4\pi} \frac{U_\lambda(k)^2 \left[ C_{j0, \lambda 0}^{j'0} \right]^2}{B j(j+1) - Bj'(j'+1) - \omega (k) }
\end{equation}
 The perturbative result can find a diagrammatic interpretation in terms of the `sunset' Feynman diagram shown in Fig.~\ref{sunset}, where the dashed line corresponds to a phonon excitation and the solid line represents the molecular state. Furthermore, the loop extends over the angular momentum variables,  momentum $k$ and frequency $\omega$. The integration over the latter leads to Eq.~\eqref{DEJM}. Since the molecular translational motion is frozen,  the molecular line in $k$ space is contracted to a single point. At the vertices, angular momentum is conserved, and within perturbation theory the incoming energy of the molecule is given by its bare energy $B j (j+1)$.
 
There are two types of processes that contribute to Eq.~\eqref{DEJM}: the ones leaving the rotational state intact, $j'=j$, and the ones changing the rotational states virtually, $j' \neq j$. Both types of  processes {can be} accompanied by the change in the quantum number $m$.

\begin{figure}[b]
\centering
\includegraphics[width=0.5\textwidth]{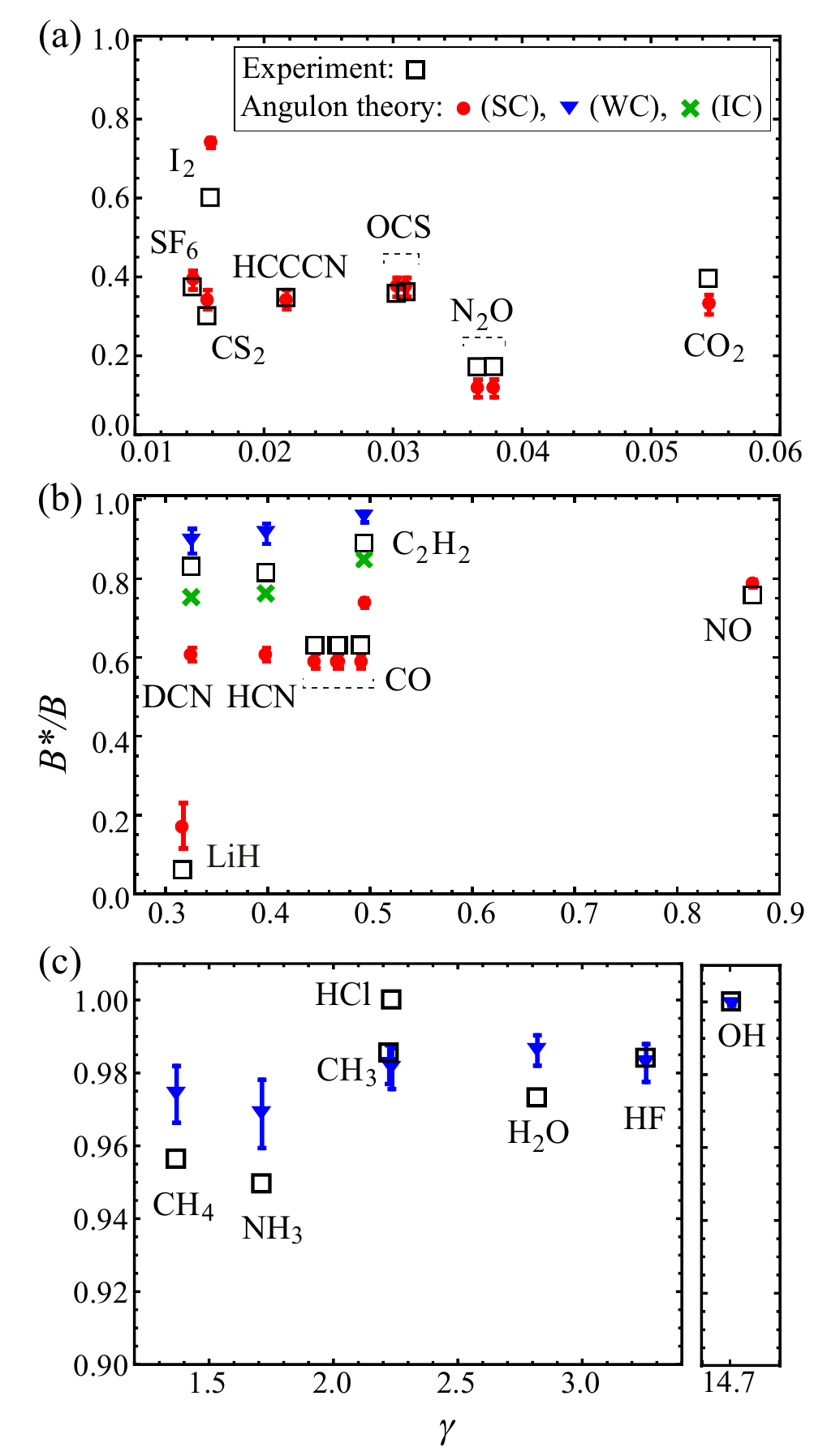}
\caption{{Renormalization of the molecular rotational constants in superfluid helium, $B^\ast/B$, as a function of the dimensionless coupling parameter, $\gamma$. Most of the heavy molecules belong to the strong-coupling regime, $\gamma<1$, while light molecules to the weak-coupling regime, $\gamma>1$.} Experimental data (empty squares) is compared with the angulon theory in the strong coupling regime (red circles), and weak coupling regime (blue triangles). The intermediate-coupling interpolation is shown by green crosses. Adapted with permission from Ref.~\cite{LemeshkoDroplets16}.}
\label{fig:BBstar}
\end{figure}

In the regime where the molecule-phonon interactions are substantially smaller than the rotational constant, the processes with $j' \neq j$ will be gapped, i.e. the virtual transitions between the levels will be suppressed by a large value of the denominator. In such a case, the energy shift will be dominated by the $j$-preserving processes:
\begin{equation}
\label{DEJM2}
	\Delta E_{jm} \approx - \sum_{k \lambda} \frac{2\lambda +1}{4\pi} \frac{U_\lambda(k)^2 \left[ C_{j0, \lambda 0}^{j0} \right]^2}{ \omega (k) }
\end{equation}
Furthermore, {due to the selection rule $j+\lambda+j = \text{even}$, imposed by the Clebsch-Gordan coefficient~\cite{VarshalovichAngMom}, only even values of $\lambda$ will contribute to Eq.~\eqref{DEJM2}. In most molecules, the interaction will be dominated by the lowest-order anisotropic term $U_2(k)$, thus giving:}
\begin{equation}
\label{DEJM3}
	\Delta E_{jm} \approx - \frac{5}{4\pi} \left[ C_{j0, 2 0}^{j0} \right]^2 \sum_{k}  \frac{U_2(k)^2}{ \omega (k) }
\end{equation}
We see that all the $j$-dependence in the equation above is incapsulated in the Clebsch-Gordan coefficient $C_{j0, 2 0}^{j0}$. Due to this dependence, the levels with different $j$ acquire different shifts, resulting in the  rotational constant renormalization, observed in several experiments on molecules in helium droplets~\cite{ToenniesAngChem04}. For example, within this approximation, for the transition $j=0 \to j=1$ which is most frequently addressed in experiments, the change in the transition energy is:
\begin{equation}
\label{DEJM4}
	h \Delta \nu_{0\to1} \approx - \frac{1}{2\pi}  \sum_{k}  \frac{U_2(k)^2}{ \omega (k) }
\end{equation}

{It is worth mentioning, that the notion of rotational constant renormalization is an approximation used to describe the lowest-energy part of the molecular rotational spectrum, such as the splitting between the $j=0$ and $j=1$ levels. In general, the energies of higher rotational states deviate from the rigid-rotor distribution~\cite{ToenniesAngChem04}.}

Recently, it has been shown that such a simple perturbative approach suffices to describe rotational constant renormalization of light molecules in $^4$He~\cite{LemeshkoDroplets16, YuliaPhysics17}. Fig.~\ref{fig:BBstar}(c) compares the results of the weak-coupling theory (blue triangles) with experiment (empty squares) {-- one can see that an agreement within 2\% is achieved}. For heavy and medium-mass species, shown in panels (a) and (b), respectively, a different, strong-coupling approach is required, see Sec.~\ref{sec:SlowRot}.

As a peculiar fact, the rotational constant renormalization for light molecules in helium represents an analogue of the Lamb shift in quantum electrodynamics~\cite{ScullyZubairy}. There, atomic states with different angular momenta acquire different shifts due to the virtual excitations of photons from the vacuum state. Similarly, in the case of a rotating molecule the virtual excitations of phonons in the superfluid lead to a state-dependent renormalization.

Furthermore,  second-order perturbation theory is a valid approximation to describe experiments on ultracold molecules coupled to a weakly-interacting BEC. In the following Sec.~\ref{sec:Angulon1}, we discuss this scenario in more detail and show that such shifts can be detected in modern experiments on ultracold quantum gases.

\subsection{Nonperturbative analysis in the weak-coupling regime}
\label{sec:Angulon1}

 {The second-order perturbation theory described in the previous section takes into account only one single-phonon excitation,  cf. Fig.~\ref{sunset}. However, even if we assume that the molecule-bath interactions are weak enough such that {simultaneous}   two- and three-phonon excitations can be neglected, multiple consecutive single-phonon excitations can still take place. In order to take  such processes into account, in this section we approach the angulon Hamiltonian~\eqref{Hamil1} from a variational perspective.} As we show below,  there exists a correspondence between the variational and diagrammatic approaches which allows one to get a deeper insight into the angulon properties. For instance, in addition to the ground-state properties accessible by standard variational approaches, it will be possible to recover the  entire absorption spectrum of the system  by making use of this correspondence. For now, however, let us start with the simplest possible variational ansatz for the many-body quantum state which is based on an expansion in single bath excitations:
\begin{equation}
\label{VarFunc}
	\vert \psi \rangle = Z_{LM}^{1/2} \ket{0} \ket{L M}+ \sum_{\substack{k \lambda \mu \\ j m}} \beta_{k \lambda j} C_{jm, \lambda \mu}^{L M} \bed_{k \lambda \mu} \ket{0} \ket{jm},
\end{equation}
with $Z_{LM}^{1/2}$ and $\beta_{k \lambda j}$ the variational parameters{, where $L$ labels the total angular momentum of the system and $M$ its projection on the laboratory-frame $Z$-axis.} Here the first ket in each term, $\ket{0}$, represents the vacuum of phonons, i.e.\ the BEC state, while the second ket refers to the molecular state. In Eq.~\eqref{VarFunc} the first term corresponds to the non-interacting state of the system where no phonons are excited and hence the total angular momentum of the system, as given by the quantum numbers $L$ and $M$, belongs to the molecule. Due to interactions between the rotating molecule and the bosons, this state is perturbed and phonons are excited from the vacuum. This is taken into account by the second term where the excitations of phonons with {momentum} $k$ and angular momentum $\ket{\lambda, \mu}$ are accompanied by the change of the molecular state to $\ket{j,m}$. This process has to conserve the total angular momentum of the system, which is incorporated explicitly by the Clebsch-Gordan coefficient $C_{jm, \lambda \mu}^{L M}$. Furthermore, unless an external field breaking the spherical symmetry is applied, all the relevant physical properties will be independent of the quantum number $M$, which we will omit hereafter. 

 The variational state defined in Eq.~\eqref{VarFunc} represents the `dressing' of the bare molecular state by fluctuations in its environment. Although its state is modified by the many-body field, the molecule retains some of its bare character and it becomes a `quasiparticle' with molecule-like properties. This quasiparticle has been termed `angulon' \cite{SchmidtLem15}. 

A question which arises is how similar the angulon is to its bare counterpart -- the unperturbed molecule. One measure for this `particle-likeness' of the angulon is given by the so-called quasiparticle weight, $Z_{LM}$. It can be determined from the coefficients of Eq.~\eqref{VarFunc} which obey the following  normalisation condition:
\begin{equation}
\label{VarFuncNorm}
	Z_{LM}{=} 1- \sum_{k \lambda j} |\beta_{k \lambda j}|^2.
\end{equation}
{The quasiparticle weight $Z_{LM}$ gives the overlap of the many-body state (molecule plus bath), $\ket{\psi}$, with the  eigenstate of the system in the absence of molecule-environment interactions, $\ket{0}\ket{LM}$, where $\ket{LM}$ reflects the state of an isolated molecule. In the regime of $Z_{LM} \to 1$,  the angulon turns into a free molecule, while in the limit of $Z_{LM} \to 0$ the molecule becomes so perturbed by fluctuations in the environment that even the quasiparticle picture (and therefore the notion of the angulon), breaks down. In what follows, we explore both of these regimes.}

Already at the level of the variational state \eqref{VarFunc} one can clearly see the difference in the treatment of angulons compared to polarons.  Unlike the perturbative variational states used in polarons~\cite{ChevyPRA06,Rath2013,Li2014,Shchadilova2016,Ashida2017}, where translational degrees of freedom are coupled, constructing the states of the form~\eqref{VarFunc} inevitably involves angular momentum algebra. For instance, including two-phonon excitations would require a $6j$-symbol, three-phonon excitations a $9j$-symbol, and so on. This problem becomes extremely challenging in the limit of strong interactions where multiple phonon excitations contribute significantly. In Sec.~\ref{sec:SLtransfo}, we introduce a canonical transformation, which allows to drastically simplify this aspect of the angulon problem.

For now, let us use Eq.~\eqref{VarFunc} to define a variational state and minimize the energy, $E = \bra{\psi}H\ket{\psi}/\langle \psi \vert \psi \rangle$, with respect to the parameters $Z_{LM}^{1/2}$ and $\beta_{k \lambda j}$. This is equivalent to minimizing the functional:
\begin{equation}
\label{MinE}
F = \bra{\psi}H - E \ket{\psi}
\end{equation}
Note that during the variational procedure, the variables, $Z_{LM}^{1/2}$ and $\beta_{k \lambda j}$, and their conjugates, $(Z_{LM}^{1/2})^\ast$ and $\beta_{k \lambda j}^\ast$, have to be treated  as independent variables; moreover, it has to be emphasized that the normalization condition \eqref{VarFuncNorm}  is to be applied only after the variation has been performed.

After minimization, we arrive at the following system of equations:
\begin{align}
\label{VarEqs1}
\frac{\partial F}{\partial (Z_{LM}^{1/2})^\ast} &= \left[B L(L+1) - E \right] Z_{LM}^{1/2} + \sum_{k \lambda j} (-1)^\lambda \sqrt{\frac{2\lambda+1}{4 \pi}} U_\lambda(k) C_{L0, \lambda 0}^{j,0} \beta_{k \lambda j} {\,\overset{!}{=}\,}0 \\
\label{VarEqs2}
\frac{\partial F}{\partial \beta_{k \lambda j}^\ast} &= \left[B j(j+1) - E + \omega (k) \right] \beta_{k \lambda j} + (-1)^\lambda \sqrt{\frac{2\lambda+1}{4 \pi}} U_\lambda(k) C_{L0, \lambda 0}^{j,0} Z_{LM}^{1/2} {\,\overset{!}{=}\,} 0
\end{align}
where we used the symmetry properties of the Clebsch-Gordan coefficients~\cite{VarshalovichAngMom}.
{As one can notice,} if one substitutes $\beta_{k \lambda j}$ from Eq.~\eqref{VarEqs2} into Eq.~\eqref{VarEqs1}, $Z_{LM}^{1/2}$ cancels, resulting in an equation where both sides depend on the energy: 
\begin{equation}
\label{Dyson1}
E = B L(L+1) - \Sigma_L(E),
\end{equation}
where
\begin{equation}
\label{Selfenergy}
\Sigma_L (E) =  \sum_{k \lambda j} \frac{2\lambda+1}{4\pi} \frac{U_\lambda (k)^2 \left[C_{L0, \lambda0}^{j0} \right]^2 }{B j(j+1) - E+ \omega (k)}.
\end{equation}
One can see that Eqs.~\eqref{Dyson1} and \eqref{Selfenergy} almost exactly coincide with the second-order perturbation theory expansion, Eq.~\eqref{DEJM}, up to one subtle difference: in the denominator,  the energy of the free molecule, $B L(L+1)$, is replaced by $E$ from the left-hand side.   In technical terms, the appearance of the energy in the denominator of Eq.~\eqref{Selfenergy} originates in the self-consistent normalization condition for the angulon wavefunction Eq.~\eqref{VarFunc}. In contrast,  second order perturbation theory does not account for the normalization of the perturbative wavefunction which {renders it less accurate compared to Eqs.~\eqref{Dyson1}--\eqref{Selfenergy}}.

In this present case, the energy has to be found self-consistently, as a solution of the equation:
\begin{equation}
\label{GFct}
	[G^0_L(E)]^{-1} - \Sigma_L(E) = 0
\end{equation}
where
\begin{equation}
\label{GFct0}
G^0_L(E) = \frac{1}{BL (L+1)-E}
\end{equation}
is the so-called `free Green's function' of the molecule, and $\Sigma_L(E)$ is the so-called `self-energy'  arising due to the dressing by the phonon field.  The specific form of Eq.~\eqref{GFct} allows to connect to a diagrammatic approach to the problem. This is reflected in the fact that  Eq.~\eqref{GFct}  is equivalent to solving for the poles in energy of the so-called Dyson equation~\cite{Mahan90} for the total angulon Green's function $G_L^\text{ang} (E)$:
\begin{equation}
\label{Dyson2}
	G_L^\text{ang} (E)= G_L^0(E) +G_L^0(E)~\Sigma_L(E)~G_L^\text{ang}(E)
\end{equation}
 where
\begin{equation}
\label{GFctInt}
G^\text{ang}_L(E) = \frac{1}{BL (L+1)-E - \Sigma_L(E) }
\end{equation}
Thus, Eq.~\eqref{GFct} gives the poles of the interacting Green's function which correspond to the  angulon's eigenenergies.

The equivalence between the variational equations~{\eqref{VarEqs1}--\eqref{VarEqs2}} and the Green's function formalism of Eqs.~\eqref{Dyson1}--\eqref{GFctInt} paves the way to the diagrammatic approach to the angulon problem. The Feynman diagrams contributing to the single-phonon expansion \eqref{VarFunc} are shown in Fig.~\ref{feynman}. These diagrams show a graphical representation of our earlier statement that the angulon is a quantum rotor dressed by a quantum many-body field. One can see that the difference between the diagrams of Fig.~\ref{feynman} and the ones of the second-order perturbation theory, Fig.~\ref{sunset}, is the bold line on the right-hand-side of self energy. This corresponds to summing over any number of sequential single-phonon excitations, which renders the resulting solution non-perturbative. Algebraically, the  resulting resummation of infinitely many diagrams originates  from the presence of energy, $E$, on both sides of Eq.~\eqref{Dyson1}.

\begin{figure}[t]
  \centering
 \includegraphics[width=0.5\linewidth]{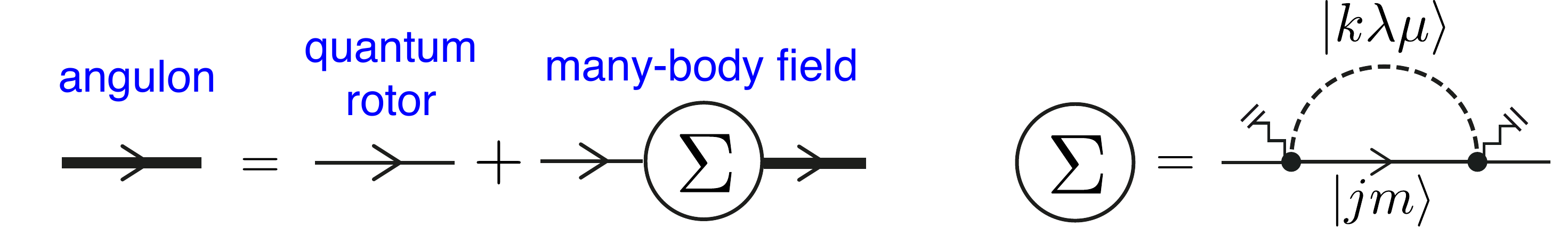}
  \caption{\label{feynman}  Representation of the Dyson equation~\eqref{Dyson2} in terms of Feynman diagrams.  }
 \end{figure}
 
{Using Eqs.~\eqref{GFctInt} and~\eqref{Selfenergy}  we are now able to calculate the angulon's Green's function, and thereby gain  access not only to the ground-state properties but also to the entire excitation spectrum of the system.} The latter is encompassed by the spectral function~\cite{AltlandSimons},
 \begin{equation}
\label{SpecFunc}
 \mathcal A_L (E)= \text{Im}[G^\text{ang}_L(E+i \varepsilon)],
 \end{equation}
which in addition provides insight  into the quasiparticle properties of the angulon. {Here $\varepsilon$ is an infinitely small positive number, $\varepsilon \to +0$.} The calculation of Eq.~\eqref{SpecFunc} requires knowledge of the self energy \eqref{Selfenergy} as a function of energy $E$, which can be calculated either numerically or using the analytic expression for its imaginary part~\cite{SchmidtLem15}:
\begin{equation}
 \label{ImSigma}
  \text{Im}\left[ \Sigma_L(E) \right]= \sum_{\lambda j k_0}\theta(E-Bj(j+1))\left[C_{L0, \lambda0}^{j0} \right]^2 \frac{2\lambda+1}{4} U_\lambda (k_0)^2 \vert(\partial \omega (k)/\partial k)_{k=k_0}\vert^{-1},
 \end{equation}
where $k_0$ gives the roots of $E-\omega (k)-Bj(j+1) =0$.  While the imaginary part $  \text{Im}\left[ \Sigma_L(E) \right]$ determines the inverse lifetime of the angulon, {the real part, $  \text{Re}\left[ \Sigma_L(E) \right]$,   gives the angulon's energy.} The theta-function in Eq.~\eqref{ImSigma} represents the onset of low-energy phonon bands, shown in Fig.~\ref{Stark}(d) (see below). Similar bands have been previously  observed in experiments with molecules in helium nanodroplets~\cite{Hartmann2002}  as well as for translationally moving impurities in a Bose-Einstein condensate of ultracold atoms \cite{Shchadilova2016,Jorgensen2016}.  

The properties of the angulon, such as its spectrum given by Eq.~\eqref{SpecFunc}, depend on the specific realization of the system. The latter are parametrized, for instance, by the dispersion relation $\omega (k)$ and the particular choice of interaction potentials. In the following we focus on one specific   example  for the angulon spectral function, which has been studied in  Ref.~\cite{SchmidtLem15}. There we considered a  superfluid with contact interactions, $ V_\text{bb} (\mathbf{q}) \equiv g_\text{bb} =4\pi a_{\text{bb}}/m$, parametrized by the boson-boson scattering length $a_{\text{bb}} >0$~\cite{Pitaevskii2016}. This resulted in the Bogoliubov dispersion relation, Eq.~\eqref{Bogwk}, having the form $\omega (k)=\sqrt{\epsilon (k)(\epsilon (k)+2 g_{\text{bb}} n)}$. 

Real atom-molecule potentials comprise multiple terms in the spherical Harmonic expansion, Eq.~\eqref{Vr}, and can be obtained using numerical quantum chemistry calculations~\cite{SzalewiczIRPC08, StoneBook13}. In the angulon Hamiltonian, Eq.~\eqref{Hamil1}, each of these terms results in a phonon-mediated coupling between the bare rotational states, and will impose its own  selection rules on the angular momentum exchange between the impurity and the bath. In order to understand the general behavior of the system, however, it is convenient to start from considering only the contributions from the leading terms $\lambda=0,1$.   In Ref.~\cite{SchmidtLem15} we assumed that   the shapes of the latter interaction potentials, {Eq.~\eqref{Vr},} were given by $V_\lambda (r) = u_\lambda f_\lambda(r)$, which are characterized by the magnitudes $u_0 = 1.75 \, u_1 = 218 B$ and  Gaussian form-factors $f_\lambda(r) = (2\pi)^{-3/2} e^{-r^2/(2r_\lambda^2)}$, with $r_0 = r_1 = 1.5\, (m B)^{-1/2}$.  This is a model potential, which does not quantitatively reproduce any particular atom-molecule system, however, does describe a typical magnitude and range of such two-body interactions. {For example, for a molecule with $B = 2 \pi \hbar \times 1$~GHz~$\approx 0.03$~cm$^{-1}$ immersed in superfluid $^4$He, the parameters correspond to the anisotropic part of the interaction $u_1 \approx 4$~cm$^{-1}$ and the range $r_1 \approx 16$~\AA.} For convenience, here and below we adapt dimensionless units, with energy measured in $B$, distances in $(m B)^{-1/2}$, and, as before, $\hbar \equiv 1$.

\begin{figure*}[t]
  \centering
 \includegraphics[width=\linewidth]{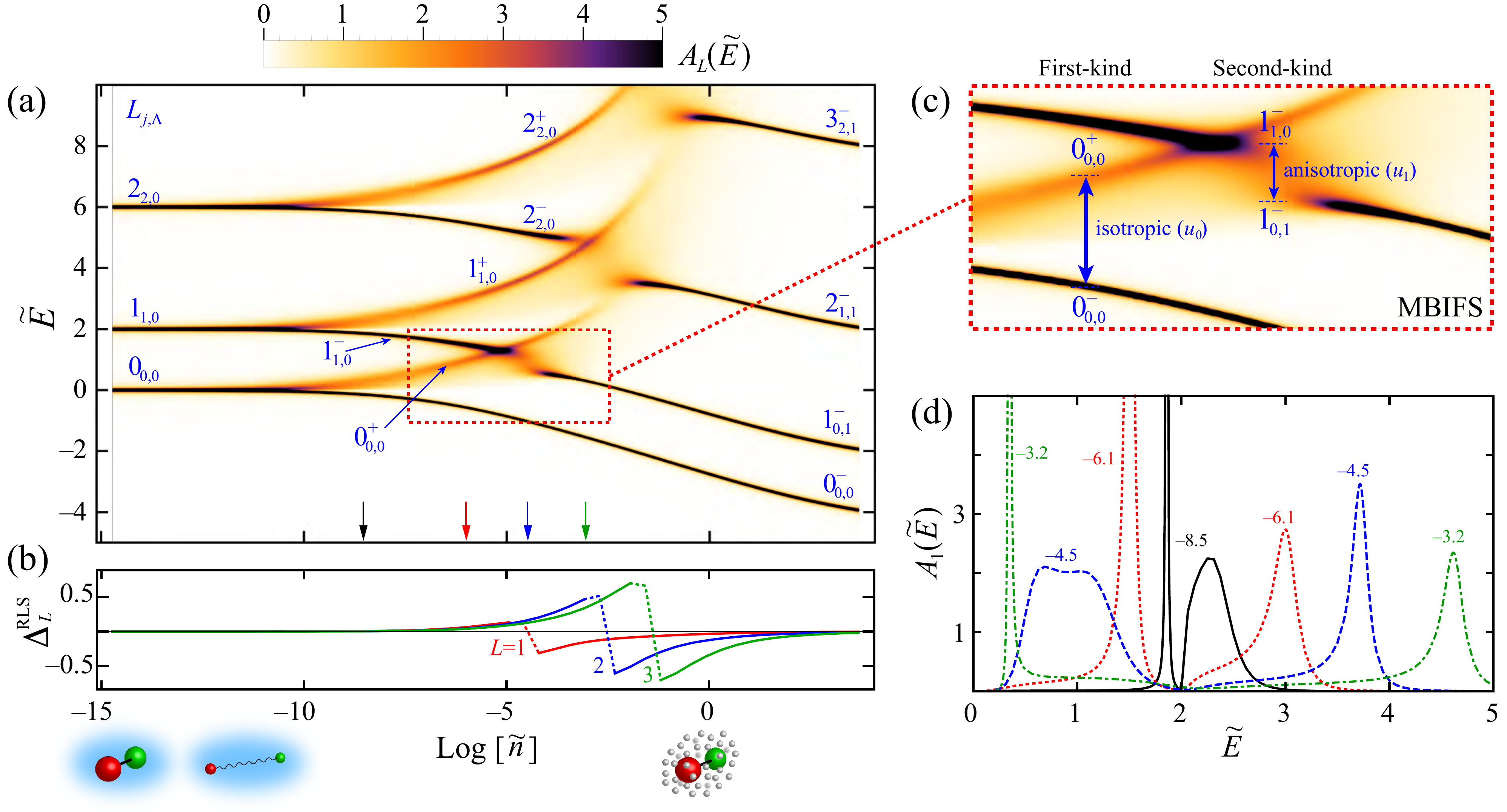}
  \caption{\label{Stark} (a) The angulon spectral function, $A_L (\tilde{E})$, as a function of the dimensionless superfluid density, $\tilde{n}=n (m B)^{-3/2}$,  {and dimensionless energy $\tilde E= E/B$. As indicated by the pictograms in the bottom, the} left side of the plot corresponds to the low density regime which is  realized for instance with ultracold {ground-state and weakly-bound} molecules interacting with a BEC of atoms. The high-density regime (right hand side of the plot) can in turn be realized with rotating molecules immersed in {superfluid helium. The states are labeled as $L_{j \Lambda}$, where $L$ is the total angular momentum of the system, and $j$ and $\Lambda$ are the angular momenta of the molecule and the bath, respectively; $+$ and $-$ are used to distinguish the states with the same  $L_{j \Lambda}$.} The arrows indicate the densities highlighted in panel (d). (b)~Differential rotational Lamb shift for the lowest nonzero-$L$ states  as given by $\tilde{\Delta}^\text{RLS}_L = (E_L - E_0)/B - L(L+1)$. (c)~Zoom-in illustrating the Many-Body-Induced Fine Structure (MBIFS) of the first kind, $L_{L,0} \to \{L_{L, 0}^-, L_{L, 0}^+\}$, and of the second kind, $L_{L,0}^- \to L_{L-1,1}^-$. (d) Spectroscopic signatures of the MBIFS for the $L=1$ state. The numbers indicate the corresponding values of $\text{Log}[\tilde{n}]$. Sharp peaks reflect long-lived angulon excitations while broad spectral features correspond to the incoherent excitation of phonons.  Adapted with permission from Ref.~\cite{SchmidtLem15}.}
 \end{figure*}

Fig.~\ref{Stark}(a) shows the resulting angulon spectral function,  $A_L(\tilde{E}) \equiv \mathcal{A}_L(\tilde{E}) B$,  with $\tilde{E} = E/B$. While the vertical axis denotes the energy $\tilde{E}$, the horizontal axis  gives the dimensionless superfluid density, $\tilde{n}=n (m B)^{-3/2}$. Thus, the left-hand side of the plot corresponds to a weakly-interacting system (such as an ultracold molecule in a BEC), while the right-hand side corresponds to  the strongly-interacting regime (e.g. a molecule in superfluid helium).

The angulon is an eigenstate of the total angular momentum $L$ of the system, which is, besides $M$, the only good quantum number. However, as often done in spectroscopy~\cite{LevebvreBrionField2}, we can introduce approximate quantum numbers: $j$, the angular momentum of the molecule, and $\Lambda$, the angular momentum of the bosons. As a result, the angulon states in Fig.~\ref{Stark}(a) are labeled as $L_{j,\Lambda}$.
Sharp peaks in the angulon spectrum (dark shade) correspond to long-lived angulon states which reproduces the spectrum of a free rotor at small densities $\tilde n$.  Furthermore the magnitude of the peaks, see Fig.~\ref{Stark}(d), determines the quasiparticle weight $Z_{LM}$ of the corresponding angulon state. The fact that the angulon closely resembles a free rotor at small densities is reflected in the fact that its quasiparticle weight approaches $Z_{LM} \to 1$ in this regime. However, once the medium becomes denser, a few peculiar effects occur (for more details see Ref.~\cite{SchmidtLem15}).

\begin{enumerate}
\item
\text{Effects due to isotropic interactions:}

\begin{itemize}
\item
\textit{Polaron shift.} As previously observed for structureless impurities~\cite{LandauPolaron,LandauPekarJETP48,AppelPolarons,Devreese13,Rath2013,Shashi2014,Shchadilova2014,Grusdt2014}, the spherically-symmetric part of the molecule-atom potential leads to the uniform lowering of all the energy levels with increasing density.

\item
\textit{Many-body-induced fine structure (MBIFS) of the first kind}, also see Fig.~\ref{Stark}(c). One can see that at intermediate densities, every $L$-level is split into a doublet. One can understand this feature as a splitting between the states $\ket{j=L , \text{no phonons}}$ and $\ket{j=L , \text{one phonon with } \lambda=0}$, coupled by the interaction potential $U_0 (k)$, which increases with the density $n$.  

\end{itemize}

\item
\text{Effects due to anisotropic interactions:}

\begin{itemize}
\item
\textit{Rotational Lamb shift (RLS)}. Due to the $\lambda=1$ term in the interaction potential, there appears a non-uniform shift, whose magnitude depends on the angular momentum $L$. Such a shift leads to the renormalization of rotational constants for molecules in superfluid helium droplets~\cite{ToenniesAngChem04}, and has also been discussed in Sec.~\ref{sec:2dOrder}. Fig.~\ref{Stark}(b) shows the rotational Lamb shift, defined as $\tilde{\Delta}^\text{RLS}_L = (E_L - E_0)/B - L(L+1)$, for a few of the lowest angular momentum states.

\item
\textit{MBIFS of the second kind}, also see Fig.~\ref{Stark}(c). There occur splittings of the angulon lines, which correspond to a resonant transfer of one quantum of angular momentum  from the rotor to the many-body  bath, $L^-_{L,0} \to L_{L-1, 1}^-$. It happens when the energy of the $L^-_{L,0}$ state is lowered by the interactions such that it merges with the phonon continuum of the angulon state with  $L-1$. Under this condition the density of states for momentum changing collisions is enhanced, leading to the splitting of the state.  Due to the enhanced phonon-molecule scattering, close to the splitting, the quasi particle weight of the angulon is largely suppressed and the quasiparticle picture breaks down.

\end{itemize}

\end{enumerate}

   \begin{figure}[t]
  \centering
    \includegraphics[width=0.5\linewidth]{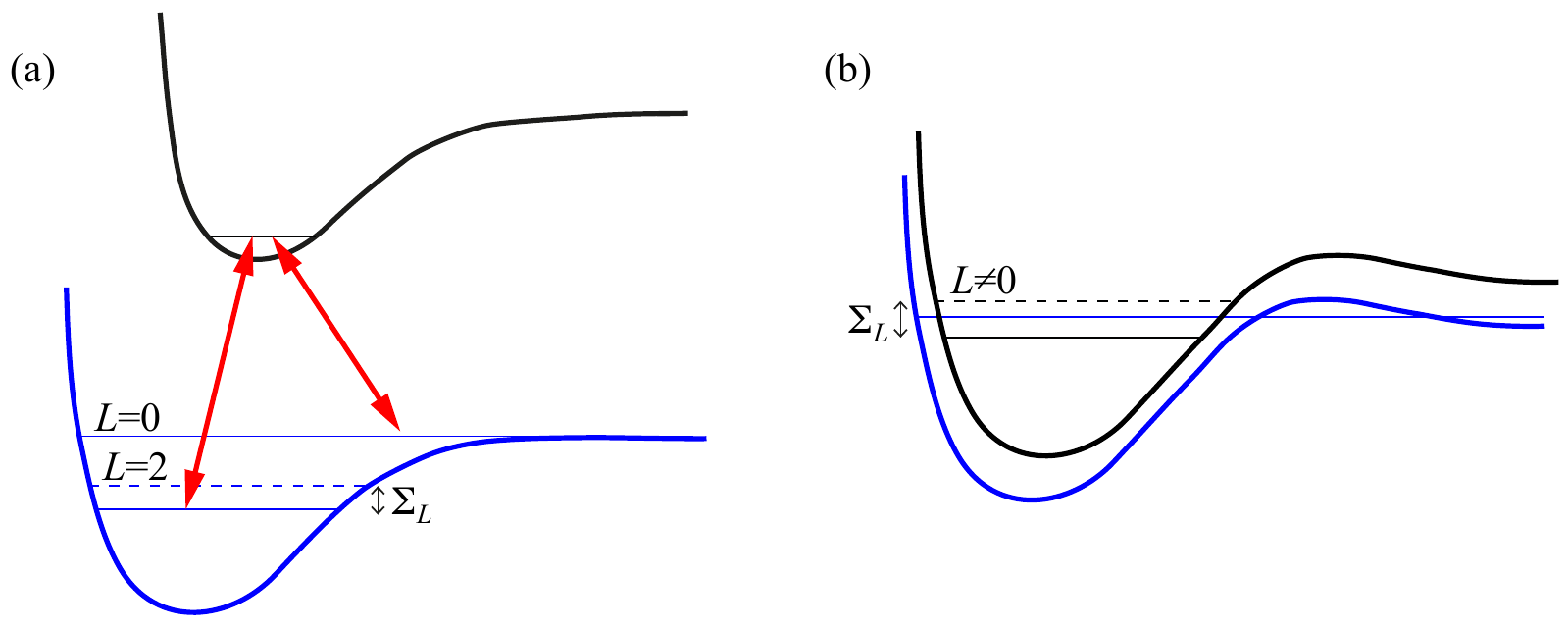}
  \caption{\label{PA} Possible schemes to detect  the angulon self-energy $\Sigma_L$ using (a)~photoassociation spectroscopy~\cite{UlmanisCR12} and (b)~shift of $p$- and $d$-wave Feshbach resonances~\cite{KohlerRMP06}.  Adapted from Ref.~\cite{SchmidtLem16}.}
 \end{figure}

The properties of the angulons, as given by the spectrum of Fig.~\ref{Stark}, could be observed  experimentally both with molecules trapped in weakly-interacting BEC's~\cite{Bikash16, Pitaevskii2016}, as well as strongly-interacting superfluids, such as helium droplets~\cite{LemeshkoDroplets16, YuliaPhysics17, ToenniesAngChem04}. For instance, the RLS can be measured by the relative shift of the rotational states of a diatomic molecule, while the MBIFS of the first and second kind can be revealed as broadenings and splittings of the rotational transition lines.  The effects are expected to be most pronounced for molecules which possess a small rotational constant $B$, such as in experiments involving molecules in highly-excited vibrational states. In the context of ultracold gases, the latter can be studied using {weakly-bound molecules~\cite{LemFriPRArapid09, LemFriPRL09, LemFriJPCA10} created by} photoassociation spectroscopy~\cite{UlmanisCR12},  or by measuring  nonzero angular momentum Feshbach resonances~\cite{KohlerRMP06},  as schematically illustrated in Fig.~\ref{PA}. {Depending on the bath density, the energies of the bound molecular states will shift, and so will the positions of the continuum-to-bound transitions.} An alternative possibility is measuring the angulon self-energy as a shift of the microwave lines in the spectra of weakly bound molecules~\cite{MarkPRA07, LemFriPRArapid09, LemFriJPCA10, LemFriPRL09}, prepared using one of these techniques. 

For molecules in superfluid helium the interactions and the bath properties cannot be tuned as easily as in ultracold gases. However, the   range of chemical species amenable to trapping is essentially unlimited~\cite{ToenniesAngChem04}, which paves the way to studying angulon physics in a broad range of parameters. We note that an effect very similar to the MBIFS of the second kind has been recently observed in the spectrum of CH$_3$ trapped in helium droplets~\cite{MorrisonJPCA13}. There, the helium environment seems to induce an (otherwise forbidden) $|\Delta K| = 3$ transition, which may find an interpretation as an  angular momentum transfer to rotons in the superfluid. However, a detailed calculation is required to confirm this interpretation of the data.

 \subsection{The canonical transformation}
 \label{sec:SLtransfo}
 
In the previous section we have seen that even   the simplest many-particle processes, such as single-phonon excitations, can drastically modify the rotational spectrum of an impurity. The next step is to consider the case where the impurity-bath interactions are strong enough to excite multiple phonons at the same time. Here we are, however, facing  a major challenge: if we were to construct a variational state analogous to Eq.~\eqref{VarFunc}, but involving two-, three-, or four-phonon excitations, we would need to make use of the $6j$, $9j$, and $12j$  symbols~\cite{VarshalovichAngMom} in order to assure the conservation of total angular momentum. In such a case, how would one approach the problem in the limit of strong molecule-bath interactions where \textit{infinitely} many phonons are excited and the angular momentum  algebra becomes intractable?

It can be demonstrated~\cite{SchmidtLem16} that there is a way around this problem, if one makes use of the following canonical transformation:
\begin{equation}
\label{Transformation}
	  \hat{S} = e^{- i \hat\phi \otimes \hat \Lambda_z} e^{- i \hat\theta  \otimes \hat\Lambda_y} e^{- i \hat\gamma  \otimes\hat \Lambda_z} 
\end{equation}
Here $(\hat\phi, \hat\theta, \hat\gamma)$ are the angle operators which act  in the  Hilbert space of the rotor, and
\begin{equation}
\label{Lambda}
	 \hat {\vec\Lambda}=\sum_{k\lambda\mu\nu}\bed_{k\lambda\mu}\boldsymbol\sigma^{\lambda}_{\mu\nu}\be_{k\lambda \nu}
\end{equation}
is the total angular momentum operator of the phonons, acting in their  Hilbert space.  The vector $\boldsymbol\sigma^{\lambda}$ is composed  of matrices   fulfilling the angular momentum algebra in the representation of angular momentum $\lambda$.
Thus, the  transformation operator of Eq.~\eqref{Transformation} uses  the total angular momentum of the bath as a generator of rotation, which transfers the environment degrees of freedom into the frame co-rotating along with the quantum rotor, as schematically illustrated in Fig.~\ref{transf}.

\begin{figure}[b]
  \centering
    \includegraphics[width=0.5\linewidth]{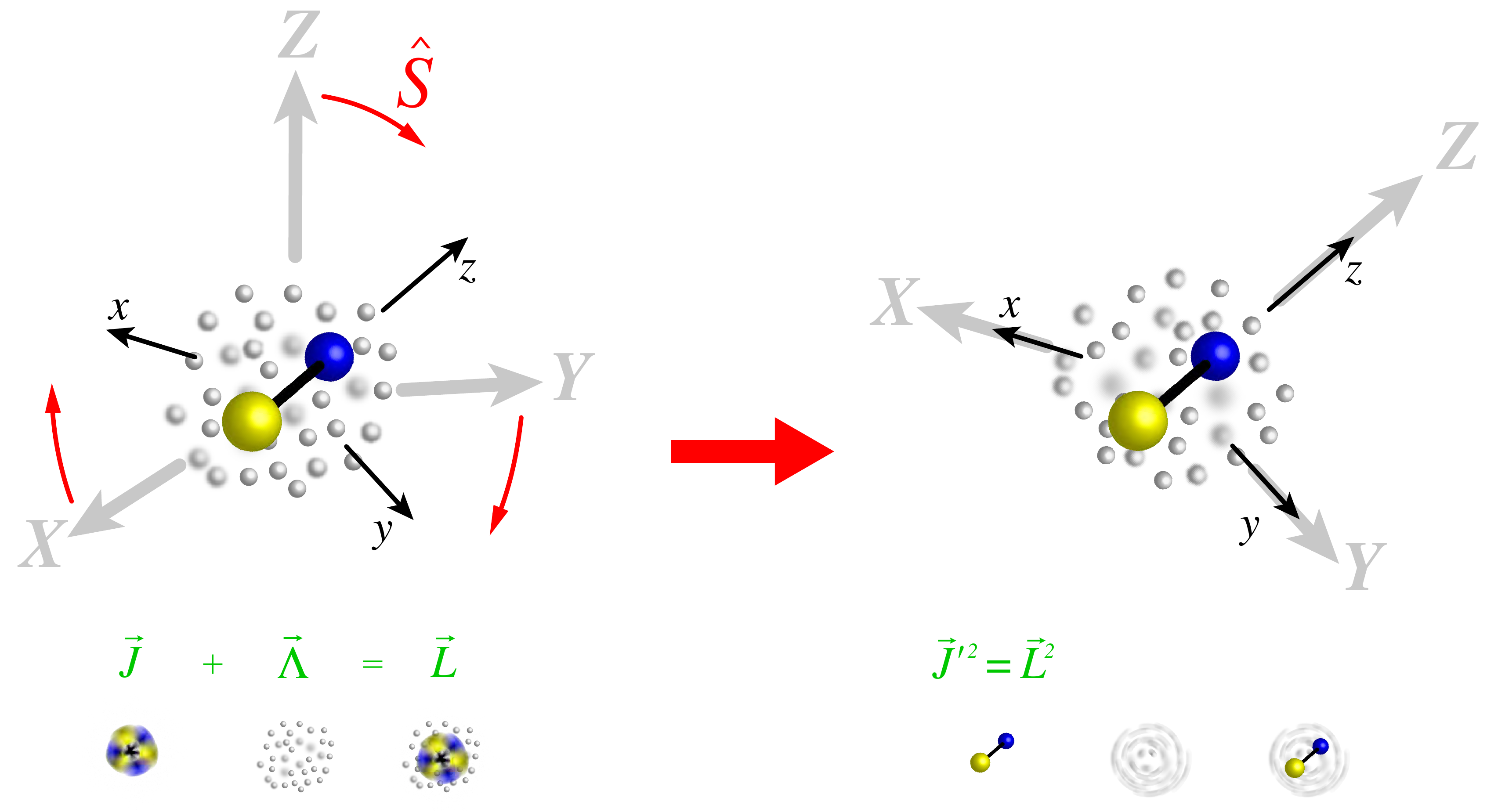}
  \caption{\label{transf}  Action of the canonical transformation, Eq. (\ref{Transformation}), on the angulon Hamiltonian,  Eq.~\eqref{Hamil1}. Left: in the laboratory frame, $(X, Y, Z)$, the combination of the molecular angular momentum, $\mathbf{J}$ and the bath angular momentum, $\boldsymbol{\Lambda}$ gives the total angular momentum of the system, $\mathbf{L}$. Right: after the transformation, the bath degrees of freedom are transferred to the rotating frame of the molecule, $(x, y, z)$. As a  consequence, the molecular angular momentum in the transformed space coincides with the total angular momentum of the system in the laboratory frame. Adapted from Ref.~\cite{SchmidtLem16}.}
 \end{figure}

The transformation~\eqref{Transformation} brings the Hamiltonian of Eq.~(\ref{Hamil1}) to the following form:
\begin{equation}
\label{transH}
\hat{\mathcal{H}} \equiv \hat S^{-1} \hat H \hat S= B (\hat{\mathbf{J}}' - \hat{\mathbf{\Lambda}})^2 + \sum_{k\lambda\mu}\omega (k) \bed_{k\lambda\mu}\be_{k\lambda\mu} + \sum_{k\lambda} V_\lambda(k) \left[\bed_{k\lambda0}+\be_{k\lambda0}\right]
\end{equation}
Here $V_\lambda(k)=U_\lambda(k) \sqrt{(2\lambda+1)/(4\pi)}$  and $\hat{\mathbf{J}}'$ is the `anomalous' angular momentum operator acting in the rotating frame of the impurity, discussed in Sec.~\ref{sec:symtops} and Appendix~\ref{sec:appendixAngular}.  The details of the derivation can be found in Ref.~\cite{SchmidtLem16}.

There are a few key features of the transformed Hamiltonian, Eq.~\eqref{transH}:

\begin{enumerate}

 \item
 \label{2lab}
\textit{$\hat{\mathcal{H}}$ does not contain the impurity coordinates $(\hat \theta,\hat \phi)$, which removes the intractable angular momentum algebra from the problem.} In particular, the spherical harmonic operators, $\hat Y_{\lambda \mu} (\hat \theta,\hat \phi)$, which couple three-dimensional angular momenta in Eq.~(\ref{Hamil1}), are now replaced by the term $\hat{\mathbf{J}}' \cdot \hat{\mathbf{\Lambda}}$, which couples only the angular momenta  \textit{projections}. Such terms occur in various problems of  physics, e.g.\ the ones involving spin-orbit and spin-spin interactions, and do not lead to $3nj$-symbols in the matrix elements of the Hamiltonian.

\item
 \textit{$\hat{\mathcal{H}}$ is explicitly expressed through the constant of motion -- the total angular momentum of the system.} It can be shown~\cite{SchmidtLem16} that the set of eigenvalues of the $\hat{\mathbf{J}}'^2$ operator in the transformed frame exactly coincides with the spectrum of the total angular momentum operator, $\hat{\mathbf{L}}^2$. 

\item
\textit{$\hat{\mathcal{H}}$ can be solved exactly in the limit of a slowly rotating impurity, $B\to 0$}. As we show in Sec.~\ref{sec:SlowRot} below, this allows to construct variational states involving an \textit{infinite} number of phonon excitations and use it to describe effective rotational constants of heavy molecules in $^4$He. Moreover, this allows to construct variational solutions which are based on the expansion around an intrinsically non-perturbative state.

\end{enumerate}

\subsection{The limit of a slowly rotating impurity}
\label{sec:SlowRot}

  \begin{figure}[b]
  \centering
    \includegraphics[width=0.4\linewidth]{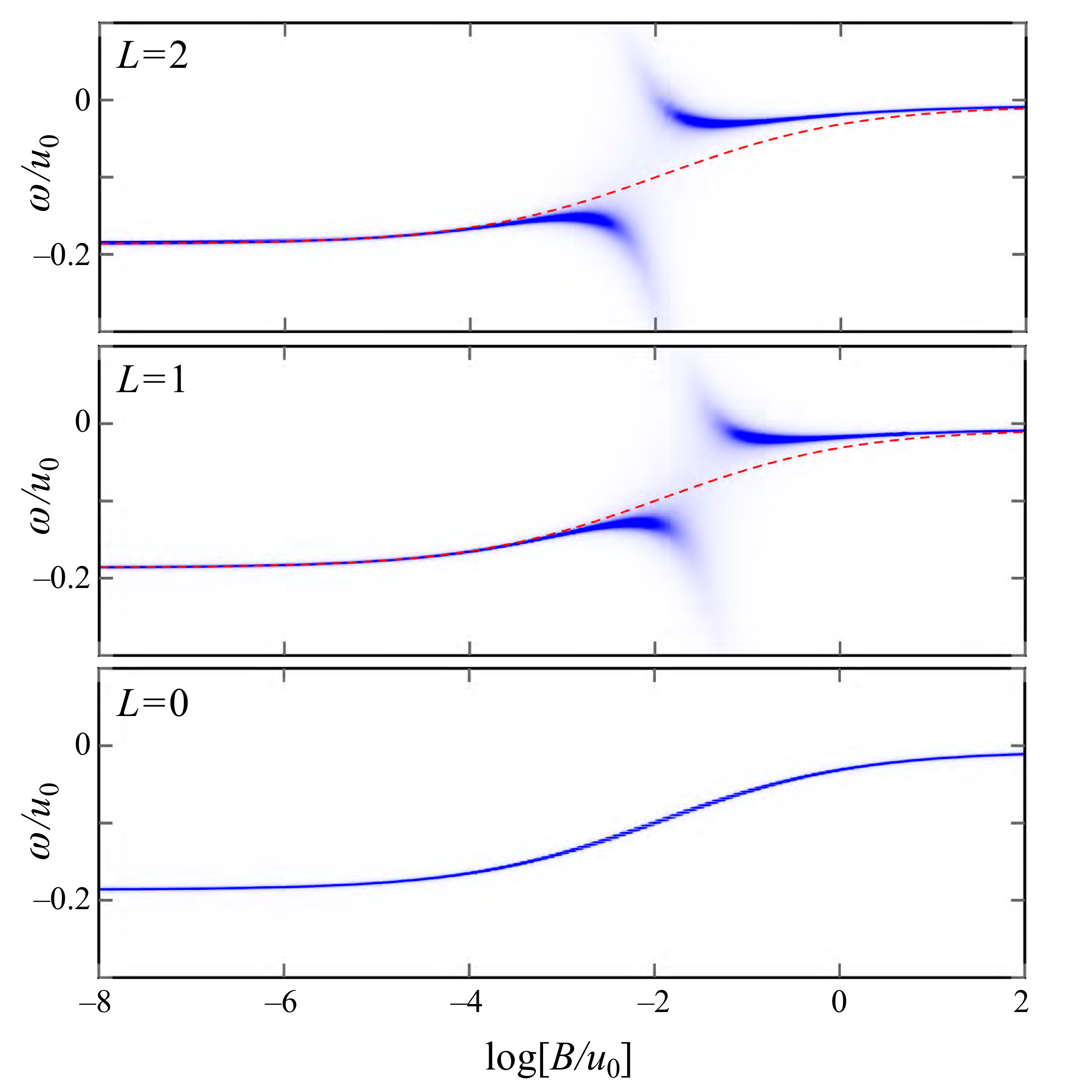}
  \caption{\label{energies} Change of the angulon spectral function, $A_L (\omega)$, where $\omega=E-B L(L+1)$, with the rotational constant $B$, for three lowest total angular momentum states. In contrast to Fig.~\ref{Stark} we work here in units of the isotropic interaction parameter $u_0$. As in our previous result  the $L>0$ states show an instability in the spectrum. The red dashed line shows the deformation energy given by the second term in Eq.~(\ref{Uhbos}), which is independent of $L$.  Adapted from Ref.~\cite{SchmidtLem16}.}
 \end{figure}
 
In the limit of a slowly-rotating impurity, $B \to 0$, the Hamiltonian (\ref{transH}) can be diagonalized exactly by an additional canonical transformation: 
\begin{equation}
\label{Htilde}
 \hat{\mathscr{H}} = \hat{U}^{-1} \hat{\mathcal{H}} \hat{U}
\end{equation}
where
\begin{equation}
\label{Utransf}
	\hat{U} = \exp \left[ \sum_{k \lambda} \frac{V_\lambda (k)}{\omega (k)} \left( \be_{k \lambda 0} - \bed_{k \lambda 0}  \right) \right]
\end{equation}

This transformation represents a coherent-state  shift of the boson operators,
\begin{align}
\label{Ushift}
	 \hat{U}^{-1} \bed_{k \lambda 0}   \hat{U} &= \bed_{k \lambda 0}   - \frac{V_\lambda (k)}{\omega (k)}\\
	 \hat{U}^{-1} \be_{k \lambda 0}    \hat{U} &= \be_{k \lambda 0}  - \frac{V_\lambda (k)}{\omega (k)}
\end{align}
which replaces the boson part of Eq.~\eqref{transH} by:
\begin{equation}
\label{Uhbos}
 	\hat H_\text{bos} = \sum_{k\lambda\mu}\omega (k) \bed_{k\lambda\mu}\be_{k\lambda\mu} - \sum_{k \lambda}  \frac{V^2_\lambda (k)}{\omega (k)}
 \end{equation}
 Since the dispersion relation $\omega (k) $ is non-negative, the ground state of the bosons is given by the phonon vacuum, $\ket{0}$. The second term of Eq.~\eqref{Uhbos} corresponds to the deformation energy of the condensate, sometimes referred to as the `chemical potential' of the impurity~\cite{ToenniesAngChem04}.

In other words, in the limit of $B \to 0$ the ground state of (\ref{transH}) represents a coherent state of bosons, $\hat U \ket{0}$, which already involves a macroscopic deformation of the condensate, i.e. an \textit{infinite} number of phonon excitations. Note that constructing such  a state starting from the original Hamiltonian \eqref{Hamil1} would be an extremely challenging task due to the angular momentum algebra involved.

The transformation~\eqref{Transformation} allows to study the regime of a slowly rotating molecule. For instance, in Ref.~\cite{SchmidtLem16} we considered a variational state  based on single  phonon excitations on top of the double transformed Hamiltonian~\eqref{Htilde} to reveal the instabilities in the angulon spectrum. As an example, Fig.~\ref{energies} shows the angulon spectral function depending on the rotational constant $B$, for a few lowest rotational states. The emerging instability is qualitatively similar to the MBIFS of the  second kind, shown in Fig.~\ref{Stark} and discussed in Sec.~\ref{sec:Angulon1} above.

Finally, the strong-coupling angulon theory allows to explain rotational constant renormalization of heavy and medium-mass molecules in helium droplets. For a slowly-rotating molecule, one can assume that the bosonic state, derived in the limit of $B\to0$ above, does not change upon molecular rotation. In a way, it can be thought of as a microscopic formulation of the `nonsuperfluid helium shell,' rotating along with the molecule~\cite{ToenniesAngChem04, SzalewiczIRPC08}. On the other hand, the effective molecular angular momentum, given by the first term of Eq.~\eqref{transH}, is determined by the difference between the total angular momentum of the system, $ \hat{\mathbf{L}} \equiv \hat{\mathbf{J}}'$, and the angular momentum of superfluid excitations, $\hat {\vec\Lambda}$. In such a way, the energy of a state with a given \textit{total} angular momentum $L$ is lower in the presence of a superfluid ($\hat {\vec\Lambda} \neq 0$) compared to a free molecule ($\hat {\vec\Lambda} = 0$), which leads to effective renormalization of the rotational constant.

In the strong-coupling regime, $\hat{\mathbf{\Lambda}}$ in Eq.~\eqref{transH} can be replaced by its expectation value, $\langle  \hat{\mathbf{\Lambda}}^2 \rangle^{1/2}$, where
\begin{equation}
\label{ExpLambda}
	  \langle  \hat{\mathbf{\Lambda}}^2 \rangle \equiv \bra{\psi_{LM}} \hat{\mathbf{\Lambda}}^2 \ket{\psi_{LM}} = \sum_{k \lambda} \lambda(\lambda+1) \frac{V^2_\lambda (k)}{\omega^2_{k}} 
\end{equation}

The results of the strong-coupling angulon theory are shown in Fig.~\ref{fig:BBstar} by red circles. From panels (a) and (b) one can see that a good agreement with experiment is achieved for most heavy and medium-mass molecules. We refer the reader to Ref.~\cite{LemeshkoDroplets16} for a detailed description of the strong-coupling theory, as well as of the intermediate-coupling interpolation, shown in Fig.~\ref{fig:BBstar} by green crosses.

\section{Conclusions and outlook}
\label{sec:conclusions}

The aim of this tutorial was to familiarize the reader with a novel perspective on interactions of molecules with a many-particle environment --  namely the one based on the notion of quasiparticles. A variety of quasiparticles has been introduced in the past in order  to describe point-like impurities, such as electrons, ultracold atoms, or single spins interacting with their respective environments. The rotational motion of molecules gives rise to a novel mechanism of impurity interactions with a bath due to the  coupling of internal and external degrees of freedom. This stems from the peculiar properties of quantum rotations such as non-Abelianity and a discrete spectrum of eigenvalues. From quantum mechanics textbooks we know that the addition of even a few angular momenta (such as photon absorption/emission by a single molecule) can already call for involved approaches. The situation clearly does not become easier if the molecular rotation occurs in the presence of a many-particle environment, where the number of angular momenta to add is in principle infinite.

We demonstrated  that a natural  way to treat a rotating molecule interacting with an environment is by introducing a new quasiparticle -- the angulon -- and investigating its properties. Along with purely theoretical derivations, we  provided an outlook  for the application of the angulon concept to molecules in BEC's as well as in superfluid helium droplets. 
For instance, we have shown that the angulon theory -- even in its most basic formulation -- allows to explain  rotational constant renormalization observed for molecules in helium droplets.  On the other hand, it predicts various phenomena, which do not occur in isolated atoms and molecules, nor in any other impurity problems. A particular example of the latter  is the angulon instability, accompanied by the resonant transfer of one quantum of angular momentum between the molecule and the many-body bath. Such instabilities are fundamentally different from vortex instabilities (see Ref.~\cite{SchmidtLem16} for a detailed comparison). Furthermore, it seems that their signatures have been recently observed in spectra of CH$_3$ molecules inside superfluid helium droplets~\cite{MorrisonJPCA13}, which, however, calls for a detailed theoretical calculation. {Among other novel phenomena featured by angulons is the angular localization of molecular impurities in the presence of a bosonic bath~\cite{Li16}.} On the other hand, nowadays rotating impurities can be prepared experimentally in perfectly controllable settings, based on ultracold molecules immersed into a Bose or Fermi gas~\cite{JinYeCRev12, KreStwFrieColdMol, LemKreDoyKais13}, which opens up a prospect for a detailed study of the angulon physics.  

Most importantly, the angulon theory provides a common language to describe  molecular impurity problems arising in different contexts. For example, although the physics of superfluid helium and dilute BEC's is substantially different, the quasiparticle picture makes the similarities and distinctions of the two settings apparent.  Moreover, the present theory has already been extended to account for \textit{ab initio} potential energy surfaces~\cite{Bikash16}, which opens up a prospect to employ the techniques of quantum chemistry in the context of quantum impurity problems. As an example, in Ref.~\cite{Bikash16} it has been shown that for a CN$^-$ ion interacting with a BEC of Sr and Rb~\cite{Bikash16}, the {rotational Lamb shifts} predicted in Ref.~\cite{SchmidtLem15} can be observed  under realistic experimental conditions. 
In the future, the presented theory can be extended to account for rotational, vibrational, and electronic structure of more complex molecules~\cite{StoneBook13}, and external electromagnetic and crystalline fields~\cite{Redchenko16, Yakaboylu16, Shepperson16, LemKreDoyKais13}, which is expected to further enrich the observed phenomena. {The question of engineering intermolecular interactions and bound states mediated by a field of ultracold atoms also represents a substantial interest~\cite{BissbortPRL13, ZhouPRA11, HerreraPRA11, HerreraPRL13, BennettPRL13, Lemeshko2013, Otterbach2014, LemFrontPhys13, LemeshkoPRA11Optical, LemFri11OpticalLong}.}  

While this tutorial entirely focused on molecules, the quasiparticle approach to the redistribution of angular momentum in many-particle systems extends far beyond molecular physics. For example, the notions of rotation and orbital angular momentum can be used to calculate the properties of nuclei~\cite{RoweWoodNuclearModels}, high-resolution atomic spectra~\cite{DereviankoRMP11}, or electronic structure of defect centers in solids~\cite{MazeNJP11}. In condensed matter physics, the orbital angular momentum of excited electronic states is often coupled to the lattice phonons, and such a redistribution of angular momentum is involved, e.g., in the ultrafast  demagnetization of ferromagnetic thin films~\cite{StammNatMat07, StammPRB10, TowsPRL15, TsatsoulisPRB16, FahnleJSNM17}. In ultracold gases, single Rydberg excitations (which can carry angular momentum) are    perturbed by phonons in the surrounding BEC~\cite{BalewskiNature13}. 
Finally, the angulon impurity problem  can be used as a building block of a general theory describing the redistribution of orbital angular momentum in quantum many-body systems. This paves the way to apply the techniques described here to other outstanding problems of chemical and condensed matter physics.

\section{Acknowledgements}

We are grateful to Giacomo Bighin, Igor Cherepanov, Eugene Demler, Gary Douberly, Bretislav Friedrich, Rytis Jursenas, Johan Mentink, Elena Redchenko, Stephan Schlemmer, Henrik Stapelfeldt, Sandro Stringari, and Andrey Vilesov for insightful discussions. The work was supported by the NSF through a grant for the Institute for Theoretical Atomic, Molecular, and Optical Physics at Harvard University and Smithsonian Astrophysical Observatory.

\appendix

\section{Angular momentum operators}
\label{sec:appendixAngular}

In the laboratory frame, $(X, Y, Z)$, the angular momentum operators obey the following commutation relations:
\begin{equation}
\label{Jcomm}
 [\hat{J}_X, \hat{J}_Y] = i \hat{J}_Z;  \hspace{1cm}  [\hat{J}_Y, \hat{J}_Z] = i \hat{J}_X;  \hspace{1cm}  [\hat{J}_Z, \hat{J}_X] = i \hat{J}_Y
\end{equation}
which can be  rewritten as follows:
\begin{equation}
\label{Jcomm2}
 [\hat{J}_i, \hat{J}_j] = i \epsilon_{ijk} \hat{J}_k
\end{equation}
Here the subscripts $i,j,k$ refer to  $X, Y, Z$ and $\epsilon_{ijk}$ is the {so-called Levi-Civita symbol: $\epsilon_{ijk}=1$ for every even permutation of $X,Y,Z$, such as $Z, X, Y$ or $Y, Z, X$; $\epsilon_{ijk}=-1$ for every odd permutation, such as $X, Z, Y$ or $Y,X,Z$; $\epsilon_{ijk}=0$ if any of the two  indices coincide, e.g.\ $X, X, Z$ or $X, Y, Y$.}

On the other hand, it turns out that in the molecular frame, $(x, y, z)$, the angular momentum operators obey \textit{anomalous} commutation relations~\cite{BiedenharnAngMom, BernathBook}:
\begin{equation}
\label{Jcomm2p}
 [\hat{J}_i', \hat{J}_j'] = -  i \epsilon_{ijk} \hat{J}_k'
\end{equation}

The components of laboratory-frame and molecular-frame operators, however, commute with the square of the angular momentum operator, as well as with one another:
\begin{equation}
\label{J2comm}
 [\hat{J}_{i}, \mathbf{\hat{J}^2}] =  [\hat{J}_{i}', \mathbf{\hat{J}^2}] =  [\hat{J}_{i}, \hat{J}_{k}']  = 0,
\end{equation}
{for all $i$ and $k$.}

We can define raising and lowering operators both in the space-fixed frame:
\begin{align}
\label{J0}
	 \hat{J}_0 &= \hat{J}_Z \\
 \label{Jplus}
	 \hat{J}_{+1} &= -\frac{1}{\sqrt{2}} \left(\hat{J}_X + i\hat{J}_Y \right)\\
\label{Jminus}
	 \hat{J}_{-1} &= \frac{1}{\sqrt{2}} \left(\hat{J}_X - i\hat{J}_Y \right)
\end{align}

and similarly in the molecular frame:
\begin{align}
\label{J0}
	 \hat{J}_0' &= \hat{J}_z' \\
 \label{Jplus}
	 \hat{J}_{+1}' &= -\frac{1}{\sqrt{2}} \left(\hat{J}_x' + i\hat{J}_y' \right)\\
\label{Jminus}
	 \hat{J}_{-1}' &= \frac{1}{\sqrt{2}} \left(\hat{J}_x' - i\hat{J}_y' \right)
\end{align}
Here we use the convention of Varshalovich~\cite{VarshalovichAngMom}. Note, however, that  the definitions of raising and lowering operators used in other sources {(e.g.\ the book of Sakurai~\cite{SakuraiQM})} can differ.

Then the action of these operators on a symmetric-top state can be described as:
\begin{align}
\label{JiKet}
 	\hat{J}_k  \vert j, m, n \rangle &=  \sqrt{j(j+1)} C_{j, m; 1, k}^{j, m+k}  \vert j, m+k, n \rangle\\
\label{JiPrimeKet}
	\hat{J}'_k  \vert j, m, n \rangle &= (-1)^{k} \sqrt{j(j+1)} C_{j, n; 1, -k}^{j, n-k}  \vert j, m, n-k \rangle
\end{align}

{Finally, let us present the commutation relations between the angular momentum operators and Wigner $\hat D$-matrices, which depend on molecular angle operators. For the laboratory-frame components of angular momentum they read:
\begin{equation}
\label{JiComm}
 	\left[\hat{J}_k, \hat D^{\lambda}_{\mu \nu}  (\hat{\phi}, \hat{\theta}, \hat{\gamma}) \right]  = (-1)^{k+1} \sqrt{\lambda(\lambda+1)} C_{\lambda, \mu; 1,-k}^{\lambda, \mu-k} \hat D^{\lambda}_{\mu-k, \nu}  (\hat{\phi}, \hat{\theta},  \hat{\gamma})
\end{equation}
\begin{equation}
\label{JiCommAst}
 	\left[\hat{J}_k, \hat D^{\lambda \ast}_{\mu \nu}  (\hat{\phi}, \hat{\theta}, \hat{\gamma}) \right]  =  \sqrt{\lambda(\lambda+1)} C_{\lambda, \mu; 1, k}^{\lambda, \mu+k} \hat D^{\lambda \ast}_{\mu+k, \nu}  (\hat{\phi}, \hat{\theta},  \hat{\gamma})
\end{equation}}

{For the molecular-frame components we have:
\begin{equation}
\label{JiPrimeComm}
 	\left[\hat{J}'_k, \hat D^{\lambda}_{\mu \nu}  (\hat{\phi}, \hat{\theta}, \hat{\gamma}) \right]  = - \sqrt{\lambda(\lambda+1)} C_{\lambda, \nu; 1, k}^{\lambda, \nu+k} \hat D^{\lambda}_{\mu, \nu + k}  (\hat{\phi}, \hat{\theta}, \hat{\gamma})
\end{equation}
\begin{equation}
\label{JiPrimeCommAst}
 	\left[\hat{J}'_k, \hat D^{\lambda \ast}_{\mu \nu}  (\hat{\phi}, \hat{\theta}, \hat{\gamma}) \right]  = (-1)^k \sqrt{\lambda(\lambda+1)} C_{\lambda, \nu; 1, - k}^{\lambda, \nu-k} \hat D^{\lambda}_{\mu, \nu - k}  (\hat{\phi}, \hat{\theta}, \hat{\gamma})
\end{equation}}


\clearpage

 \bibliography{ColdChemistry_Refs}

\end{document}